\documentclass[superscriptaddress,dvipsnames,nofootinbib,amsmath,amssymb,prd,notitlepage]{revtex4-1}

\usepackage{natbib} 

\usepackage{float}
\usepackage{array}

\usepackage{subfig}
\usepackage[a4paper, left=20mm,right=20mm,top=20mm,bottom=20mm]{geometry}
\usepackage[titletoc,toc,title]{appendix}

\usepackage{mathtools}
\usepackage{tensor}
\usepackage{amsmath, amsfonts,bm}
\usepackage{derivative}
\usepackage{bm}
\usepackage[utf8]{inputenc}
\usepackage[normalem]{ulem}
\usepackage{nicefrac}

\usepackage{suffix}

\usepackage{amsmath, amsfonts,bm}
\usepackage{xcolor}
\usepackage{graphicx}
\definecolor{darkerred}{rgb}{0.9, 0.17, 0.31}
\usepackage{subfig}
\usepackage{multirow}

\usepackage[colorlinks=true,hyperfootnotes=true, linkcolor=darkerred, citecolor=cyan, urlcolor=magenta]{hyperref}

\DeclareMathAlphabet{\mathup}{OT1}{\familydefault}{m}{n}

\newcommand{\be}{\begin{equation}} 
\newcommand{\ee}{\end{equation}}

\usepackage[capitalise]{cleveref}

\usepackage{array}
\newcommand{\PreserveBackslash}[1]{\let\temp=\\#1\let\\=\temp}
\newcolumntype{C}[1]{>{\PreserveBackslash\centering}p{#1}}
\newcolumntype{R}[1]{>{\PreserveBackslash\raggedleft}p{#1}}
\newcolumntype{L}[1]{>{\PreserveBackslash\raggedright}p{#1}}
\renewcommand\arraystretch{1.5}

\begin{document}

\title{Breaking dark degeneracies in \texorpdfstring{$k$}{k}-essence models with redshift-space distortions}
\author{Rebecca Briffa}
\email{rebecca.briffa.16@um.edu.mt}
\affiliation{Institute of Space Sciences and Astronomy, University of Malta, Malta, MSD 2080}

\author{Lehel Csillag}
\email{lehel.csillag@ubbcluj.ro}
\email{lehel.csillag@unitbv.ro}
\affiliation{Department of Physics, Babeș-Bolyai University, Kogălniceanu
Street, Cluj-Napoca 400084, Romania} 
\affiliation{Faculty of Mathematics and
Computer Science, Transilvania University, Iuliu Maniu Street 50, Brașov
500091, Romania}
\author{Erik Jensko}
\email{erik.jensko@ucl.ac.uk}
\email{jenskoerik@gmail.com}
\affiliation{Department of Mathematics, University College London,
Gower Street, London WC1E 6BT, UK}

\begin{abstract}
Dynamical dark energy models are often difficult to constrain with background observations alone due to parameter degeneracies. This is especially true when the dark energy sector evolves through an effective pressureless matter phase, leading to degeneracies between the model parameters and the
matter density parameter. In this work, we show that redshift-space distortion (RSD) measurements  help break this degeneracy when combined with expansion history probes including Pantheon$^{+}$ supernovae and DESI DR2 BAO measurements. 
We demonstrate this in the ghost and dilatonic ghost condensate $k$-essence models, restricting the observational analysis to physically admissible
solutions by imposing the absence of ghosts and gradient
instabilities. Given the small scalar sound speeds in these
models, we also explicitly check the validity of the
quasi-static approximation used in the growth analysis.
Our results show that combining growth-rate data with stability requirements and standard background observations provides a fast and efficient way to constrain potentially degenerate dark sector models.

\end{abstract}

\maketitle

{\hypersetup{linkcolor=black}
\makeatletter
\def\l@subsubsection#1#2{}
\makeatother
\tableofcontents
}

\clearpage

\section{Introduction}

The standard $\Lambda$CDM model provides an excellent phenomenological description of a wide range of cosmological observations, from measurements of the cosmic microwave
background (CMB) to the evolution of large-scale structure in the Universe
\cite{SupernovaCosmologyProject:1998vns,SupernovaSearchTeam:1998fmf,
Planck:2018vyg,DES:2026fyc}. 
Despite the remarkable successes of the $\Lambda$CDM model, there are several unresolved challenges. At the fundamental level, the precise nature of the dark sector is unknown, and the value of the cosmological constant is difficult to reconcile with vacuum energy predictions from quantum field theory \cite{Weinberg:1988cp,Martin:2012bt}. Additionally, $\Lambda$CDM is faced with the long-standing Hubble tension, the discrepancy between the present-day expansion rate $H_0$ inferred from CMB observations and the value obtained from local distance measurements
\cite{Planck:2018vyg,Riess:2021jrx,H0DN:2025lyy}.

There has also been considerable interest in potential discrepancies concerning the amplitude of matter clustering,
as quantified by $S_8$ \cite{Heymans:2020gsg,Wright:2025xka,DES:2026fyc}. While the level of agreement between weak-lensing and CMB determinations depends on the dataset and analysis, these comparisons have motivated further tests of the growth of structure. More recently, DESI DR2 baryon acoustic oscillation (BAO) measurements, combined with CMB and Type Ia supernova observations, have shown a preference for a time-dependent dark energy equation of state, with a significance that depends on the supernova sample \cite{DESI:2025zgx}.
These observational challenges have motivated numerous studies of extensions to the standard cosmological model
\cite{Verde:2019ivm,Abdalla:2022yfr,CosmoVerseNetwork:2025alb}.

However, many extended dark sector models are difficult to constrain with background observations alone due to
parameter degeneracies \cite{Kunz:2007rk,Kunz:2009yx,Aviles:2011ak}. At the background level, this is particularly relevant when the dark energy sector evolves through an effective
pressureless matter phase, $w_{\rm DE}\simeq0$, allowing its contribution to the expansion history to be partly
degenerate with the matter density parameter $\Omega_m$. Such behaviour arises naturally in $k$-essence theories, in which a scalar field has a non-canonical kinetic term
\cite{Armendariz-Picon:2000nqq,Armendariz-Picon:2000ulo,
Scherrer:2004au}. Indeed, recent background analyses of these models leave
$\Omega_m$ weakly constrained and dependent on the data
combination \cite{Hussain:2024qrd,Csillag:2025gnz},
motivating the use of complementary probes sensitive
to the clustering properties of the dark sector.

Redshift-space distortions (RSD) provide such a probe through measurements of the growth-rate combination $f\sigma_8(z)$. Peculiar velocities introduce anisotropies
in the observed galaxy distribution, allowing the growth of matter perturbations to be constrained in addition
to the background expansion \cite{Hamilton:1997zq}.
Growth measurements can therefore help distinguish clustering properties of dark energy models which are otherwise degenerate at the background level \cite{Tsujikawa:2012hv}. For example, RSD measurements have been shown to
substantially reduce degeneracies between the dark
matter density and the interaction strength in
interacting dark energy models \cite{Ghedini:2024mdu}.

In this work, we compare the ability of background and growth-rate measurements to break parameter degeneracies in two concrete $k$-essence models: the ghost condensate and the dilatonic ghost condensate \cite{Arkani-Hamed:2003pdi,Scherrer:2004au,Piazza:2004df}.
To ensure perturbative stability, we impose the absence of ghosts and gradient instabilities directly in the MCMC analysis as theory priors. We then constrain our models using
RSD measurements with cosmic chronometers,
SH0ES-calibrated Pantheon$^{+}$ supernovae and
DESI DR2 BAO data.
We find that BAO measurements substantially reduce the degeneracy in the ghost condensate model, while RSD measurements are crucial for breaking the degeneracy in the more complicated dilatonic model. We also explicitly ensure the validity of the quasi-static approximation on the relevant
RSD scales, which is particularly important for $k$-essence models with small scalar propagation speeds, where the subhorizon condition alone does not justify neglecting dark
energy perturbations
\cite{Bamba:2011ih,Tsujikawa:2012hv,Sawicki:2015zya}.
Our results demonstrate that growth-rate data, combined with stability requirements and standard background observations, provide a fast and efficient way to constrain otherwise degenerate dark sector models using late-time data alone, without requiring a full Boltzmann analysis.

The paper is outlined as follows. In \cref{sec:2} we introduce the $k$-essence framework and the two models we study, giving their stability conditions, equations
of state and propagation speeds. In \cref{sec:3} we
derive the background cosmological equations and the
growth-rate equation. In \cref{sec:obs_data} we describe
the observational datasets and the implementation of
the stability conditions. The observational constraints
and model comparison are presented in
\cref{sec:results}, followed by our conclusions.

\section{Ghost condensate \texorpdfstring{$k$}{k}-essence models} \label{sec:2}
The $k$-essence theories describe a scalar field with a non-canonical
kinetic term
\cite{Armendariz-Picon:1999hyi,Armendariz-Picon:2000nqq,Armendariz-Picon:2000ulo, Lueiza-Colipi:2026bqc, Lueiza-Colipi:2026gij}. Within the $k$-essence theories, we focus on two particular models
in which higher-order kinetic terms allow stable accelerated
expansion without a scalar potential.
The \textit{ghost condensate} model provides a simple realisation
of a time-dependent scalar background with a nonzero kinetic term which approaches a cosmological constant equation of state at the condensate point
\cite{Arkani-Hamed:2003pdi}.
The second model we study is the \textit{dilatonic ghost condensate},
a string-inspired extension in which the quadratic kinetic term
has an exponential dependence on the scalar field
\cite{Piazza:2004df}. These will be given explicitly in the next part.

The action for the generic $k$-essence theories is given by 
\begin{equation} \label{k_action}
    S_{\phi} = \int \sqrt{-g} P(\phi,X) d^4 x\, ,
\end{equation}
where $X =-\nabla^{\mu}\phi\nabla_{\mu}\phi/2$ is the kinetic term. The associated energy-momentum tensor is 
\begin{equation} \label{T_phi}
    T_{\mu \nu}^{\phi} = -\frac{2}{\sqrt{-g}}\frac{\delta S_{\phi}}{\delta g^{\mu \nu}} = g_{\mu \nu} P + P_{,X} \nabla_{\mu}\phi \nabla_{\nu} \phi \, ,
\end{equation}
from which we identify the effective fluid four-velocity $u^{\mu}_{\phi}= \nabla^{\mu} \phi/\sqrt{2X}$ and obtain the energy density $\rho_{\phi} = 2X P_{,X}-P$ and pressure $p_\phi = P$ \cite{Armendariz-Picon:1999hyi}. The equation of state is then simply \begin{equation}\label{w_phi}
    w_{\phi} := \frac{p_\phi}{\rho_{\phi}} = \frac{P}{2X P_{,X}-P } \, .
\end{equation}

In order for the models to be stable on cosmological backgrounds, we require a non-negative propagation speed\footnote{Note that the higher derivative operators are relevant in the strongly coupled regime $c_s^2 \rightarrow 0$, see \cite{Arkani-Hamed:2003pdi,Babichev:2018twg} for more detailed discussions.} and the absence of ghosts.
The importance of perturbative stability is evident in
recent studies of monodromic $k$-essence, where
observational constraints were restricted to the
background evolution because of the difficulties
associated with phantom crossing
\cite{Goldstein:2025epp}.
These conditions  take the form \cite{Garriga:1999vw}
\begin{align} \label{cs2}
    c_s^2 &:= \frac{P_{,X}}{\rho_{\phi,X}} = \frac{P_{,X}}{P_{,X}+2X P_{,XX}} \geq 0 \, , \\ 
    Q_s &= P_{,X} + 2 X P_{,XX}  > 0  \label{ghost} \, , 
\end{align} 
ensuring the absence of Laplacian instabilities and ghost degrees of freedom respectively. Additionally, we require a non-negative energy density $\rho_{\phi} \geq 0$ as a background viability condition. Nonetheless, we will see that for the models studied here, imposing both $c_s^2 \geq 0$ and $Q_s>0$ ensures $\rho_{\phi} \geq 0$, so the latter does not introduce any additional constraint.

The Einstein field equations take the form
\begin{equation}
    G_{\mu \nu} = \kappa \big(T_{\mu \nu}^{\phi} + T_{\mu \nu}\big)\, ,
\end{equation}
where $\kappa=8\pi G$, $T_{\mu \nu}^{\phi}$ is defined in \cref{T_phi} and the perfect fluid matter energy-momentum tensor is 
\begin{equation}
     T_{\mu \nu} = \left(\rho_{m} + p_{m} \right) u_{\mu} u_{\nu} +  g_{\mu \nu} p_{m} \, .
\end{equation}
The matter sector is minimally coupled and separately conserved $\nabla^{\mu}T_{\mu \nu}=0$.
The density and pressure are subject to the null energy condition $\rho_{m} + p_{m} >0$, avoiding instabilities in the uncoupled matter sector. For the pressureless dark matter models considered in this paper, this reduces to $\rho_{m} >0$. In what follows, we set $p_{m}=0$.
Lastly, the $k$-essence scalar field equations of motion are
\begin{equation}
    \left( P_{,X} g^{\mu \nu} - P_{,XX} \nabla^{\mu} \phi \nabla^{\nu} \phi\right) \nabla_{\mu} \nabla_{\nu}\phi + P_{,\phi X}\nabla^{\mu}\phi\nabla_{\mu}\phi + P_{,\phi} =0 \, .
\end{equation}

Lastly, we point out that while $k$-essence theories have also been proposed as models of a
unified dark sector \cite{Scherrer:2004au}, reproducing both dark
matter clustering and accelerated expansion places strong
restrictions on their perturbations. A non-negligible sound speed
can produce acoustic oscillations and suppress structure growth,
as demonstrated for unified fluid models
\cite{Sandvik:2002jz,Giannakis:2005kr}.
Purely kinetic models can avoid these effects by maintaining a
sufficiently small sound speed, but this can require considerable
fine-tuning \cite{Scherrer:2004au,Giannakis:2005kr}.
We therefore include a standard cold dark matter component and
treat the $k$-essence field as dark energy.
Crucially, a phase with $w_{\phi}\simeq0$ does not imply that the
scalar clusters identically to cold dark matter. Their different clustering
behaviour allows growth measurements to distinguish contributions
that are otherwise degenerate in the background expansion.

\subsection{Model~1: Ghost condensate}
The ghost condensate model provides a simple realisation of
purely kinetic $k$-essence
\cite{Arkani-Hamed:2003pdi,Scherrer:2004au}, described by
the Lagrangian
\begin{equation}
    P(X)=-X+\alpha X^2 \, ,
\end{equation}
where $\alpha$ is a positive constant. The linear
$X$ term has the opposite sign to a standard canonical scalar
field, but the quadratic term allows for stable solutions and gives rise to a ghost condensate point $P_{,X}=0$ situated at
$X=1/(2\alpha)$ \cite{Arkani-Hamed:2003pdi}
The  energy density  and equation of state are given by 
\begin{equation} \label{w_phi_g}
 \rho_\phi = X (3 X \alpha -1)  \, , \qquad w_{\phi} =  \frac{1 - \alpha X}{1 - 3 \alpha X}  \,, 
 \end{equation}
 while the sound speed and no-ghost constraints take the form
\begin{equation} \label{cs2_g}
     c_s^2 = 
    \frac{1 - 2 \alpha X}{1 - 6 \alpha X} \geq 0 \, , \qquad Q_s = 6 \alpha X - 1> 0 \, .
\end{equation}
Taken together, we have that $X \geq  1/(2\alpha)$. This explains why only positive values of $\alpha$ are considered, given that $X$ is strictly positive on cosmological backgrounds. 
The same condition is obtained by enforcing $c_s^2 \geq 0$ and $\rho_{\phi} \geq 0$ \cite{Csillag:2025gnz}.
The stability conditions in \cref{cs2_g} translate to constraints on the equation of state, $-1 \leq w <1/3$, ruling out phantom behaviour, consistent with usual null energy condition arguments
\cite{Dubovsky:2005xd}. This automatically implies subluminal propagation speeds, with $c_s^2 <1/3$. At the limiting condensate point $X = 1/(2\alpha)$, the field behaves like a cosmological constant $w_\phi=-1$ with a vanishing propagation speed $c_s^2=0$.

A comprehensive cosmological dynamical systems analysis of this model for both signs of $\alpha$ was given in \cite{Csillag:2025gnz}, where it was shown that the stable ghost condensate case $\alpha >0$ can give rise to a viable late-time cosmological evolution. In the presence of standard pressureless matter, the model admits an intermediate matter-dominated phase, in addition to a stable late-time de Sitter fixed point and an unstable early time radiation-tracking fixed point \cite{Csillag:2025gnz}. The radiation tracking behaviour and the approach to a cosmological constant equation of state are also familiar features of kinetic $k$-essence \cite{Armendariz-Picon:2000nqq,Scherrer:2004au}.

\subsection{Model~2: Dilatonic ghost condensate}

The dilatonic ghost condensate model is a string-inspired
extension of Model~1 in which the quadratic kinetic term
has an exponential dependence on the scalar field
\cite{Piazza:2004df}. The Lagrangian is given by
\begin{equation} \label{P_dil}
    P(\phi,X)=-X+\beta e^{\lambda\phi}X^2 \, ,
\end{equation}
where $\beta>0$ and $\lambda\geq0$ are constants.
This reduces to the previous model in the limit
$\lambda=0$, with $\alpha=\beta$. 
The energy density and equation of state take the form
\begin{equation} \label{w_phi_dil}
    \rho_\phi=X(3\beta Xe^{\lambda\phi}-1) \, , \qquad
    w_\phi=\frac{1}{3}
    -\frac{2}{9\beta Xe^{\lambda\phi}-3} \, ,
\end{equation}
and the sound speed and no-ghost constraints are
\begin{equation} \label{cs2_dil}
    c_s^2=\frac{1}{3}
    -\frac{2}{18\beta Xe^{\lambda\phi}-3}\geq0 \, ,
    \qquad Q_s=6\beta Xe^{\lambda\phi}-1>0 \, .
\end{equation}
Together, these stability conditions require
$Xe^{\lambda\phi}\geq1/(2\beta)$.
As in Model~1, the same condition follows from requiring
$c_s^2\geq0$ and $\rho_\phi\geq0$, and the stable branch
satisfies $-1\leq w_\phi<1/3$ and $0\leq c_s^2<1/3$, ruling out both stable phantom and superluminal behaviour \cite{Piazza:2004df}.

A detailed dynamical systems analysis was given in
\cite{Gumjudpai:2005ry}. For the uncoupled model considered
here, the non-phantom scalar-dominated attractor gives
accelerated expansion for
$0\leq\lambda<\sqrt{2/3}\simeq0.8165$. We use this requirement to motivate the prior
on $\lambda$ in the observational analysis.
For $\lambda>0$, the accelerating attractor has
$w_\phi>-1$, giving rise to a stable and dynamical late-time equation of
state \cite{Piazza:2004df,Gumjudpai:2005ry}.

A summary of the key properties of both models is given in \cref{tab:model_properties}.

\begin{table}[h]
\centering
\small

\renewcommand{\arraystretch}{1.5}
\setlength{\extrarowheight}{3pt}
\setlength{\tabcolsep}{8pt}

\begin{tabular}{|l|c|c|c|c|}
\hline
Model
& $\rho_\phi$
& $w_\phi$
& $c_s^2$
& $Q_s$
\\[2pt]
\hline

Ghost condensate
& $X(3\alpha X-1)$
& $\displaystyle \frac{\alpha X-1}{3\alpha X-1}$
& $\displaystyle \frac{2\alpha X-1}{6\alpha X-1}$
& $6\alpha X-1$
\\[5pt]
\hline

Dilatonic ghost condensate
& $X(3\beta X e^{\lambda\phi}-1)$
& $\displaystyle
  \frac{\beta X e^{\lambda\phi}-1}
       {3\beta X e^{\lambda\phi}-1}$
& $\displaystyle
  \frac{2\beta X e^{\lambda\phi}-1}
       {6\beta X e^{\lambda\phi}-1}$
& $6\beta X e^{\lambda\phi}-1$
\\[5pt]
\hline
\end{tabular}
\caption{Background and perturbative quantities for the ghost and
dilatonic ghost condensate $k$-essence models.}
\label{tab:model_properties}
\end{table}

\section{Cosmological dynamics} \label{sec:3}
We work with the spatially flat Friedmann–Lema\^{i}tre–Robertson–Walker (FLRW) metric given by
\begin{equation}
  ds^2=-dt^2+a(t)^2\delta_{ij}dx^idx^j \, ,
\end{equation}
where $a(t)$ is the scale factor.
The background cosmological equations in the presence of a single pressureless matter fluid, representing baryons and cold dark matter with $p_m=0$, are
\begin{align}
    3 H^2 &= \kappa(\rho_{m} + \rho_{\phi}) \, , \\
    2\dot{H} + 3H^2 &= - \kappa p_{\phi}\, ,
\end{align}
where $H = \dot{a}/a$. The $k$-essence energy density $\rho_{\phi}$ and pressure $p_{\phi}$ are defined for both models in the previous section, where $X=\dot{\phi}^2/2$. The background scalar field equation is
\begin{equation}
\left(P_{,X}+2X P_{,XX}\right)\ddot{\phi} + 3H P_{,X}\dot{\phi} + 2X P_{,X\phi} - P_{,\phi} = 0 \, .
\end{equation}
The separately conserved pressureless matter component has the standard evolution in terms of redshift $z$,
\begin{equation}
    \rho_m= \frac{3H_0^2\Omega_{m0}(1+z)^3}{\kappa} \,,
\end{equation}
where $H_0$ and $\Omega_{m0}$ are the present day values of the Hubble constant and density parameter respectively. We now give the explicit background equations to be solved for the ghost and dilatonic ghost condensate models. In what follows, we work in units of $\kappa=1$.

\subsection{Model~1 background equations}

We follow the derivation laid out in our previous work \cite{Csillag:2025gnz}, introducing the following dimensionless quantities\footnote{We include the same numerical factor of $3/4$ in $\alpha_{\rm phys}$ to match the previous work \cite{Csillag:2025gnz}, where the same dimensionless parameter is labelled as $B$.} 
\begin{equation}
     h(t) =\frac{H(t)}{H_0}  \, ,   \qquad u(t) = \frac{\dot{\phi}(t)}{H_0}\,, \quad \alpha_{\rm phys} = \frac{3\alpha H_0^2}{4}\, , 
     \end{equation}
where $h$ is the dimensionless Hubble factor, $u$ relates to the kinetic term and $\alpha_{\rm phys}$ is the positive dimensionless coupling parameter. 
The stability conditions can be written as
\begin{equation}
    c_s^2 = \frac{3-4 \alpha_{\rm phys} u^2}{3-12 \alpha_{\rm phys} u^2} \geq 0 \,, \qquad Q_s = 4 \alpha_{\rm phys} u^2 -1 >0
\end{equation}
which combine to give $4 \alpha_{\rm phys} u^2 \geq 3$. In redshift variables $1+z=1/a$, the Friedmann constraint takes the form 
\begin{equation}
\label{ff1}
    3h(z)^2 = 3 \Omega_{m0}(1+z)^3-\frac{1}{2} u(z)^2+  \alpha_{\rm phys}  u(z)^4 \, , 
    \end{equation}  
and the dynamical cosmological equations can be written as \cite{Csillag:2025gnz}
\begin{align}  
    \label{FF1}
    \frac{dh(z)}{dz}&=\frac{18h(z)^2-3 u(z)^2+ 2 \alpha_{\rm phys}  u(z)^4}{12\left(1+z\right)h(z)} \, , \\
\label{Phi1}
    \frac{d u(z)}{dz}&=\frac{u(z) \left(4 \alpha_{\rm phys} u(z)^2-3\right) }{(1+z) \left(4 \alpha_{\rm phys} u(z)^2-1\right)} \, .
\end{align}

To make direct comparison with $\Lambda$CDM, we evaluate the Friedmann constraint at redshift $z=0$, allowing us to trade the initial scalar field velocity 
value $u_0$ with the matter density parameter $\Omega_{m0}$. Imposing $h(0)=1$ and $u(0)=u_0$ in \cref{ff1}, we obtain the relation
\begin{equation} \label{ff2}
    1 = \Omega_{m0} - \frac{1}{6} 
    u_0^2 + \frac{\alpha_{\rm phys} }{3} u_0^4 \, ,
\end{equation}
which can be solved for $u_0$,
\begin{equation} \label{usol}
    u_0^2 =\frac{1+\sqrt{1+48\alpha_{\rm phys}(1-\Omega_{m0})}}{4\alpha_{\rm phys}} \,,
\end{equation}
where the stable branch has been taken (see \cite{Csillag:2025gnz}).
At $z=0$, the stability requirement  $4 \alpha_{\rm phys} u^2 \geq 3$ can be written as
\begin{equation} \label{model_constr1}
    0 \leq \Omega_{m0} < 1 \, , \quad \alpha_{\rm phys} \geq \frac{1}{16-16\Omega_{m0}} \, ,
\end{equation}
where we have also included bounds on $\Omega_{m}$ for stability of the matter sector. Hence, the set of free model parameters after using the solution of \cref{usol} is $\{H_0,\alpha_{\rm phys},\Omega_{m0}\}$.

For this model, if the stability conditions are satisfied at initial times $(z=0)$ it follows that they are automatically satisfied at all redshifts. This is because the stable part of the model forms a closed submanifold of the system \cite{Csillag:2025gnz}. It is therefore sufficient to implement the condition \cref{model_constr1} in the initialisation of the observational analysis.
For the dilatonic ghost model of the next section, this is not the case, so the no-ghost and no-gradient instability conditions need to be evaluated at all redshifts for posterior samples to be accepted.

Lastly, it is illustrative to point out that the lower bound on $\alpha_{\rm phys}$ corresponds to the limiting trajectory where the model reproduces the exact $\Lambda$CDM background. This is precisely the ghost condensate point of the purely kinetic $k$-essence models, namely the constant-$X$ branch with $P_{,X}=0$ \cite{Scherrer:2004au,dePutter:2007ny,Arkani-Hamed:2003pdi}. This trajectory coincides with the boundary of the stable parameter space \cite{Csillag:2025gnz}, which in our parametrisation is
\begin{equation} \label{alpha_lambda}
    \alpha_{\rm phys} = \alpha_\Lambda(\Omega_{m0}):= \frac{1}{16(1-\Omega_{m0})}\,.
\end{equation}
For these values the $k$-essence field has $\rho_\phi+p_\phi=2XP_{,X}=0$ and $w_\phi=-1$ with a vanishing propagation speed $c_s^2=0$. Consequently, this behaves like a cosmological constant at the background level and the expansion reduces to the $\Lambda$CDM form. 

\subsection{Model~2 background equations} \label{sec:back2}
For the dilatonic ghost condensate it is convenient to introduce the
dimensionless variables 
\begin{equation}
    x(t)=\frac{\dot{\phi}}{\sqrt{6}H}\,,
    \qquad
    y(t)=\frac{e^{-\lambda\phi/2}}{\sqrt{3\beta}H} \,, 
\end{equation}
where we assume the expanding branch $H>0$ and consequently $y>0$. 
We also take $\lambda \geq 0$, following the standard conventions used in the dilatonic ghost condensate literature \cite{Gumjudpai:2005ry}. The sign of $x$ labels two distinct branches: the $x<0$ branch is associated with a phantom scalar field dominated fixed point which is excluded by the stability conditions; the stable ghost-condensate point of interest is in the $x>0$ branch \cite{Gumjudpai:2005ry}. We therefore impose $x>0$ as a physical prior. Finally, we restrict $0 \leq \lambda \leq 0.817$ to ensure that the late-time fixed point gives rise to accelerating expansion, as derived in \cite{Gumjudpai:2005ry}.

In these variables the stability conditions become
\begin{equation} \label{mod2stab}
c_s^2=\frac{2x^2-y^2}{6x^2-y^2} \geq 0 \, \qquad 
    Q_s=-1+\frac{6x^2}{y^2}>0   \, .
\end{equation}
Thus the physically viable branch is then subject to $2x^2\geq y^2$, whereupon both of the above constraints are satisfied. In the redshift representation, the Friedmann constraint can be written as
\begin{equation}
   3h(z)^2 = 3\Omega_{m0}(1+z)^3 + h(z)^2 \left( \frac{9 x(z)^4}{y(z)^2}-3x(z)^2\right)\, .
\end{equation}
Evaluating the constraint at $z=0$, with $h(0)=1$ and $x=x_0$, gives the initial conditions for $y$
\begin{equation} \label{y0}
    y_0  = \frac{\sqrt{3}x_0^2}{\sqrt{1+x_0^2-\Omega_{m0}}} \,, 
\end{equation}
where the positive branch is chosen to satisfy $y>0$. 
Combining this expression with the stability condition
$c_s^2\geq0$ gives $x_0\leq\sqrt{2-2\Omega_{m0}}$, which we impose as an initial prior.
The final system of first-order equations is therefore
\begin{align}
    \frac{dh}{dz} \label{dd1}
    &= -\frac{3h\left(-x^4-y^2+x^2y^2\right)}{2(1+z)y^2}\,,
    \\ \label{dd2}
    \frac{dx}{dz}
    &= -\frac{3\left(6x^7+2x^3y^2-\sqrt{6}\lambda x^4y^2-7x^5y^2+xy^4+x^3y^4\right)}{2(1+z)y^2(6x^2-y^2)}\,,
    \\ \label{dd3}
    \frac{dy}{dz}
    &=-\frac{3x^4+3y^2-\sqrt{6}\lambda xy^2-3x^2y^2}{2(1+z)y}\,,
\end{align}
with the Friedmann constraint being used to eliminate $y_0$ in favour of $\Omega_{m0}$ from \cref{y0}. An explicit derivation of \cref{dd1,dd2,dd3} from the cosmological equations is given in  Appendix \ref{append:calc}. Note that the singularities of the system given by \cref{dd1,dd2,dd3}, such as the singular surface $6x^2-y^2=0$ or $y=0$, are avoided due to the stability conditions $2x^2> y^2$ and $y>0$. To summarise, the stability priors on the initial input parameters are
\begin{equation}
    0 < \Omega_{m0} < 1 \, , \qquad
    0 < x_0 < \sqrt{2-2\Omega_{m0}} \, .
\end{equation}
The dynamical conditions for the absence of ghosts and gradient
instabilities in \cref{mod2stab} are checked along each trajectory.
In practice, we require $y>0$, $6x^2-y^2>0$ and $2x^2-y^2>0$,
ensuring $Q_s>0$ and $c_s^2>0$, together with
$\Omega_m(z)=1+x^2-3x^4/y^2>0$, evaluated over the relevant redshifts.

The number of free functions and parameters in the final system has been reduced from five $\{h,\phi, \dot{\phi},\lambda, \beta \}$ down to four $\{h,x,y,\lambda\}$. In the MCMC analysis, the model parameters will therefore be $\{H_0,x_0,\lambda,\Omega_{m0}\}$. Lastly, note that there is no exact $\Lambda$CDM limit for this model, except when $\lambda =0$, for which the model reduces to the pure ghost condensate case.


\subsection{Growth rate equation}
To obtain the growth rate equation, we
consider scalar perturbations in the Newtonian gauge
\begin{equation}
  ds^2=-(1+2\Phi)dt^2+a(t)^2(1-2\Psi)\delta_{ij}dx^idx^j \, ,
\end{equation}
where $\Phi$ and $\Psi$ are the Newtonian metric potentials. It is well-known that the anisotropic stress vanishes in $k$-essence models \cite{Bamba:2011ih}, so we will later set the metric potentials equal to one another $\Phi=\Psi$ without loss of generality. To obtain the relevant equations, we follow the
standard treatment used in RSD analyses working in the subhorizon quasi-static regime \cite{Tsujikawa:2012hv}.

The model is minimally coupled so the conservation equations for the pressureless matter component with $p_m = \delta p_m = 0$ are
unchanged. In Fourier space these take the form
\begin{equation}
    \dot{\delta}_m + \frac{\theta_m}{a} - 3\dot{\Psi} = 0 \, , \qquad 
    \dot{\theta}_m + H\theta_m - \frac{k^2}{a}\Phi = 0 \, ,
\end{equation}
where $\delta_m=\delta\rho_m/\rho_m$ is the matter density contrast and $\theta_m$ is the velocity divergence. There is no anisotropic stress so $\Psi=\Phi$. On scales well inside the Hubble radius $k \gg aH$, provided that the scalar field perturbations give a negligible contribution to the gravitational potential on the scales of interest, the Poisson equation reduces to \cite{Bamba:2011ih,Tsujikawa:2012hv}
\begin{equation} \label{eq:Poisson}
    \frac{k^2}{a^2}\Phi \simeq -4\pi G \rho_m \delta_m \, ,
\end{equation}
which is the standard GR expression, with the restored Newton's constant for clarity.
The $k$-essence sector then only affects matter growth through the background expansion, with no modification to the effective Newton's constant.
Combining the above equations gives an equation for the growth factor 
\begin{equation}
    \ddot D + 2H\dot D - 4\pi G\rho_m D = 0 \,, 
\end{equation}
where $\delta_m(t,\mathbf{k})=D(t)\delta_m(\mathbf{k})$ in the scale-independent
quasi-static limit. Equivalently, in terms of redshift we have
\begin{equation} \label{eq:D}
    D''(z) + \left[ \frac{H'(z)}{H(z)} - \frac{1}{1+z} \right]D'(z) -  \frac{3\Omega_{m0}(1+z)}{2h^2(z)}D(z)  = 0 \,, 
\end{equation}
where primes denote derivatives with respect to $z$. Thus, the growth rate equation takes its standard form as in $\Lambda$CDM, with the $k$-essence modifications entering only through $H(z)$.

Defining the logarithmic growth rate factor
\begin{equation}
    f(z)=\frac{d\ln D}{d\ln a}
    =-(1+z)\frac{D'(z)}{D(z)} \,,
\end{equation}
we can write \cref{eq:D} as
\begin{equation} \label{eq:f}
    f^2(z) + \left[ 2-(1+z)\frac{H'(z)}{H(z)}\right]f(z)-(1+z)f'(z)
    =  \frac{3\Omega_{m0}(1+z)^3}{2h^2(z)} \,.
\end{equation}
In the numerical integration of \cref{eq:D}, we normalise $D(0)=1$ and sample $D'(0)$ independently, together with $\sigma_{8,0}$. We marginalise over these parameters, remaining agnostic about the earlier perturbation evolution and its matching to the late-time growth equation. Since we do not impose a pure growing-mode
initial condition, the allowed solutions can contain both
independent modes of \cref{eq:D}.
The RSD observables are computed as
\begin{equation}
    f\sigma_8(z) = f(z)\sigma_{8,0}\frac{D(z)}{D(0)} \, ,
\end{equation}
where $\sigma_{8}$ is the root mean square mass fluctuation amplitude smoothed over the scale $8h^{-1}$Mpc \cite{Heymans:2020gsg}, and $\sigma_{8,0}$ is the present-day value. Lastly, it is useful to introduce the derived clustering parameter $S_8 =\sigma_{8,0} \sqrt{\Omega_{m,0}/0.3}$, used to quantify the weak-lensing tension \cite{Heymans:2020gsg,Wright:2025xka,DES:2026fyc}, which will be discussed further in the following section.

It is important to point out that the validity of the quasi-static limit and the consequent subhorizon growth equation is not automatic for $k$-essence models \cite{Bamba:2011ih}. In addition to being deep inside the horizon $k\gg aH$, the contribution of the dark energy perturbations to the Poisson equation must be subdominant $|\Omega_m \delta_m| \gg |\Omega_\phi \delta_\phi|$ on the RSD scales of interest (typically $k\sim 0.02$ to $0.2h {\rm Mpc}^{-1}$). This can occur either because the scalar perturbations are pressure-supported, or because their gravitational contribution is suppressed by the background evolution $\Omega_{\phi}(1+w_{\phi})$. The former amounts to $k\gg k_s \sim a H/c_s$, which can become problematic in models where $c_s^2$ is small. Indeed, in both of our models the $k$-essence propagation speed can evolve close to zero, but because $w_{\phi}$ is close to $-1$ at these times, the dark energy perturbations remain suppressed. This conclusion was also stated for the dilatonic ghost condensate model in \cite{Tsujikawa:2012hv}. In Appendix \ref{append:QS} we  check that $k\gg k_s$ or $\Omega_\phi(1+w_{\phi})\ll 1$ over the relevant redshift ranges, ensuring the consistency of the quasi-static approximation and the use of \cref{eq:f} in our analysis.

\section{Observational data and methodology}
\label{sec:obs_data}

This section describes the observational probes used to test these $k$-essence models, the likelihood with which each probe enters the analysis, and the methodology employed for ensuring theoretical stability. The data combination comprises the Pantheon$^{+}$ sample of Type Ia supernovae (SNIa), including the SH0ES Cepheid host-distance anchors, baryon acoustic oscillation (BAO) measurements, cosmic chronometers (CC), and redshift-space distortion (RSD) growth-rate measurements.

\subsection{Datasets}
\subsubsection{Cosmic chronometers}

Cosmic chronometers constrain $H(z)$ directly. The method uses spectroscopic age estimates of passively evolving galaxies. The age difference $\Delta t$ between two such galaxies separated by $\Delta z$ yields
\begin{equation}
    H(z) = -\frac{1}{1+z}\,\frac{\Delta z}{\Delta t}\,.
\end{equation}
Because the measurement is differential, the inference does not require a fiducial cosmology, nor an integral over the expansion history, but it rests on the absolute dating of the stellar populations \cite{Jimenez:2001gg}.

We adopt the compilation of Moresco et al. \cite{Moresco:2012jh,Moresco:2015cya,Moresco:2016mzx}, together with the covariance matrix of Ref.~\cite{Moresco:2020}, which accounts for the correlations induced by astrophysical systematics. The corresponding likelihood is
\begin{equation}
    \chi^2_{\mathrm{CC}}
    =
    \bigl[\boldsymbol{H}(\Theta) - \boldsymbol{H}_{\mathrm{obs}}\bigr]^{\mathsf{T}}
    C_{\mathrm{CC}}^{-1}
    \bigl[\boldsymbol{H}(\Theta) - \boldsymbol{H}_{\mathrm{obs}}\bigr]\,,
\end{equation}
where $H_i(\Theta)=H(z_i,\Theta)$ is the theoretical Hubble parameter.

\subsubsection{Type Ia supernovae: Pantheon+}
For SNIa we use the Pantheon+ sample, including the SH0ES Cepheid host-distance anchors \cite{Scolnic:2021amr,Brout:2022vxf,Riess:2021jrx}. The catalogue contains 1701 light curves of 1550 distinct
Type Ia supernovae over $0.001\leq z\leq2.26$. Hubble-flow supernovae enter with $z_{\rm HD}>0.01$, and the Cepheid calibrators are retained. Including the SH0ES anchors in the likelihood breaks the degeneracy between the absolute magnitude $M$ and $H_0$. We denote this SH0ES-calibrated Pantheon+ sample by PN in the dataset combinations.

The distance modulus relates the apparent magnitude $m$ to the luminosity distance,
\begin{equation}
    \label{eq:dist_mod}
    \mu(z_i,\Theta) = m - M = 5\log_{10}\bigl[D_L(z_i,\Theta)\bigr] + 25\,,
\end{equation}
with
\begin{equation}
    D_L(z_i,\Theta)
    = c\,(1+z_{{\rm HEL},i})\int_0^{z_{{\rm HD},i}}\frac{\mathrm{d}z'}{H(z',\Theta)}\,.
\end{equation}
The absolute magnitude $M$ is sampled as a nuisance parameter. For a calibrator, this distance modulus is the Cepheid host modulus $\mu_i^{\rm Cepheid}$.

The supernova contribution to the likelihood is \cite{SNLS:2011lii}
\begin{equation}
    \chi^2_{\mathrm{SN}}
    = \Delta\boldsymbol{\mu}(\Theta)^{\mathsf{T}}\,
      C_{\mathrm{SN}}^{-1}\,
      \Delta\boldsymbol{\mu}(\Theta)\,,
\end{equation}
where $\Delta\mu_i = m_{B,i}^{\rm corr}-M-\mu_i^{\rm Cepheid}$ for a calibrator and $\Delta\mu_i = m_{B,i}^{\rm corr}-M-\mu(z_i,\Theta)$ for a Hubble-flow supernova, and $C_{\mathrm{SN}}$ is the Pantheon+ covariance, including statistical and systematic uncertainties.

\subsubsection{Baryon acoustic oscillations}
\label{sec:bao}
We use the DESI DR2 BAO measurements, based on three years of observations \cite{DESI:2025zgx}. These measurements span seven redshift bins over $0.1<z<4.2$ and constrain the transverse comoving distance $D_M(z)$, the Hubble distance $D_H(z)$, and their angle-averaged combination $D_V(z)$, each reported relative to the comoving sound horizon at the drag epoch, $r_d$. 

In this work, the drag-epoch sound horizon $r_d$ is left free and marginalised over, rather than fixed to the Planck value. The BAO data constrain only distance ratios such as $D_M/r_d$ and $D_H/r_d$, so this uses the redshift dependence of the acoustic scale without imposing early-Universe assumptions on the late-time expansion history. The mentioned distances entering the likelihood are
\begin{align}
    D_H(z) &= \frac{c}{H(z)}\,,\\
    D_M(z) &= (1+z)\,D_A(z)
            = c\int_0^{z}\frac{\mathrm{d}z'}{H(z')}\,,\\
    D_V(z) &= \left[z\,D_M(z)^2\,\frac{c}{H(z)}\right]^{1/3}\,.
\end{align}
The data vector is assembled from the ratios published by DESI. The bright-galaxy sample at $z_{\mathrm{eff}}=0.295$ reports $D_V/r_d$. The remaining bins, at $z_{\mathrm{eff}}=\{0.510,0.706,0.934,1.321,1.484,2.330\}$, report $D_M/r_d$ and $D_H/r_d$. The BAO contribution to the likelihood is
\begin{equation}
    \chi^2_{\mathrm{BAO}}(\Theta)
    = \Delta\boldsymbol{\mathcal{G}}(\Theta)^{\mathsf{T}}\,
      C_{\mathrm{BAO}}^{-1}\,
      \Delta\boldsymbol{\mathcal{G}}(\Theta)\,,
\end{equation}
where $\Delta\mathcal{G}_i(\Theta)=\mathcal{G}(z_i,\Theta)-\mathcal{G}_{\mathrm{obs}}(z_i)$ and $C_{\mathrm{BAO}}$ is the covariance published with the DR2 measurements.

\subsubsection{Growth-rate data}
\label{sec:rsd}

The growth-rate compilation is that of Table~II in Appendix~A of Ref.~\cite{Alestas:2022gcg}, combining redshift-space distortion and peculiar velocity measurements of $f\sigma_8$. In galaxy redshift surveys, overdensities are elongated along the line of sight on small scales and compressed on large scales, so a map built from spectroscopic redshifts is anisotropic even when the underlying density field is statistically isotropic. Surveys quote the combination $f\sigma_8(z)\equiv f(z)\,\sigma_8(z)$ rather than $f(z)$ alone, because $f\sigma_8$ is independent of the galaxy bias \cite{Kazantzidis:2018rnb,Lambiase:2018ows,Gonzalez:2016lur,Gupta:2011kw}. Converting redshifts to distances also introduces an Alcock--Paczynski distortion. Following Ref.~\cite{LeviSaid:2021yat}, we treat that contribution as negligible and do not apply an AP correction.

The theoretical prediction $f\sigma_8(z_i,\Theta)$ is obtained from Eq.~\eqref{eq:D} and Eq.~\eqref{eq:f}. With $\Delta Q_i(\Theta) = f\sigma_8(z_i,\Theta) - f\sigma_{8,\mathrm{obs}}(z_i)$, the growth likelihood is
\begin{equation}
    \chi^2_{\mathrm{RSD}}
    = \Delta\boldsymbol{Q}(\Theta)^{\mathsf{T}}\,
      C_{\mathrm{RSD}}^{-1}\,
      \Delta\boldsymbol{Q}(\Theta)\,.
\end{equation}
The covariance is diagonal except for the WiggleZ block \cite{WiggleZ:2012sek} and the correlation between the eBOSS measurements at $z=0.38$ and $z=0.51$
\cite{Alam:2020sor}.

\subsection{Stability implementation}

We impose the background and stability constraints derived in the previous sections as theoretical priors in the MCMC analysis. Agreement with background observations alone does not
ensure the absence of ghosts or gradient instabilities
\cite{Goldstein:2025epp}, so these conditions must be
imposed independently of the observational likelihood. For
Model~1, provided that the initial conditions are stable and satisfy \cref{model_constr1}, the model will remain free from ghost and gradient instabilities throughout
its evolution \cite{Csillag:2025gnz}. It is therefore sufficient to impose these
constraints on the initial input parameters. Model~2, on the
other hand, can have well-behaved initial conditions with
$c_s^2>0$ and $Q_s>0$, but develop an unstable evolution.
The stability conditions must therefore also be checked along
each background trajectory.

For Model~2, we impose the initial constraints
$0<x_0<\sqrt{2-2\Omega_{m0}}$, together with the sampling
ranges listed in \cref{tab:priors}.
The restriction on $\lambda$ selects models with an
accelerating late-time fixed point. After solving the background equations,
we evaluate the conditions
$y>0$, $6x^2-y^2>0$, $2x^2-y^2>0$ and $\Omega_m(z)>0$
on a grid of 800 points over $0\leq z\leq2.4$.
These ensure that the background remains on the regular branch
with positive matter density and satisfies both the no-ghost
and no-gradient-instability conditions over the redshift
range considered. 

Parameter points violating any of these conditions, or for
which the solver fails or returns non-finite values, are assigned
zero likelihood (i.e.\ $\log\mathcal{L}=-\infty$) and rejected
by the sampler. In particular, this excludes trajectories
that become gradient-unstable before reaching the singular
surface $Q_s=0$, where the background equations become
ill-defined.

\section{Observational results} \label{sec:results}
Parameter constraints are obtained by Markov chain Monte Carlo sampling with the public \textit{emcee} package\footnote{The code used for our analysis is available at \url{https://github.com/Rebecca-Briffa/k_essence_RSD}.} \cite{ForemanMackey:2012ig}.
The resulting chains provide the one-dimensional marginal posterior of each parameter and the joint posterior of each pair, from which the $1\sigma$ and $2\sigma$ credible regions reported below are derived.

Across all analyses we leave free the background parameters $H_0$ and $\Omega_{m,0}$, together with the PN absolute-magnitude nuisance parameter $M$. When BAO data are included we also leave free the sound-horizon scale $r_{\mathrm{d}}$. For combinations including RSD, we normalise $D(0)=1$
and sample the initial condition $D'(0)$ together with the amplitude $\sigma_{8,0}$. In addition, Model~1 includes the free parameter, $\alpha_{\mathrm{phys}}$ \cref{model_constr1}, while Model~2 includes the free parameters $x_0$ and $\lambda$, \cref{dd1,dd2}. We consider the combinations CC+PN, CC+PN+BAO, CC+PN+RSD and CC+PN+BAO+RSD. Sampling priors used for each of the models and dataset combinations are given in \cref{sec:lcdm}.

\subsection{Model~1 constraints}

\begin{figure}[htbp]
    \centering
    \includegraphics[width = \linewidth]{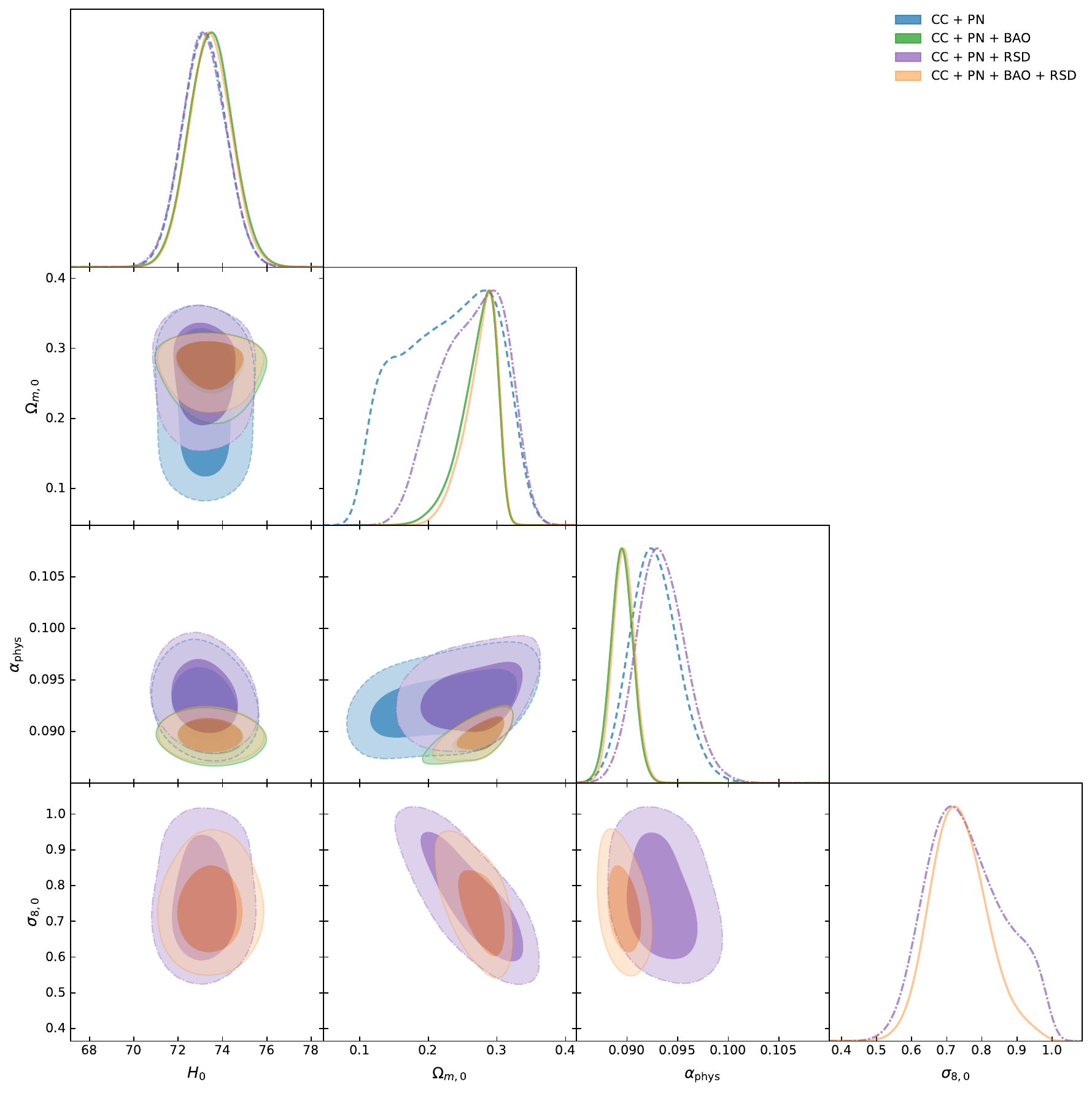}
    \caption{Confidence contours and posterior distributions for Model~1. Diagonal panels show the one-dimensional posteriors, and off-diagonal panels the corresponding $68\%$ and $95\%$ constraints, for CC+PN, CC+PN+BAO, CC+PN+RSD and CC+PN+BAO+RSD. The amplitude $\sigma_{8,0}$ is shown only for the combinations that include RSD data.}
    \label{fig:model1}
\end{figure}

The posterior distributions and parameter constraints
for Model~1, obtained by solving \cref{FF1,Phi1}, are
shown in \cref{fig:model1} and \cref{tab:model1}. The one-dimensional posteriors for $H_0$ lie almost on top of one another. Both the central value and the $1\sigma$ width change very little when BAO or RSD data are added to CC+PN, so the expansion rate today is fixed mainly by the chronometer and supernova measurements. What does change is the shape of the joint contours. In the CC+PN combination, $H_0$ remains degenerate with $\Omega_{m,0}$ where a higher matter density can be partly compensated by a shift in $H_0$ while still passing through the same distance and expansion-rate data. Once the full combination is used, that degeneracy is largely lifted and the fit gives 
$H_0=73.41^{+1}_{-0.928}\,\mathrm{km\,s^{-1}\,Mpc^{-1}}$.

The matter density parameter $\Omega_{m,0}$ is very sensitive to the choice of dataset, as is clear both from the one-dimensional posteriors and from the uncertainties in the table. With CC+PN alone the posterior for $\Omega_{m,0}$ is broad, and the contours extend over a wide range of matter densities. Adding the RSD measurements already pulls this posterior in, because the growth of structure is sensitive to how much matter is available to cluster. Replacing the growth data by BAO tightens the constraint further, since the acoustic scale limits the expansion history that $\Omega_{m,0}$ is allowed to produce. The most stringent result is obtained when all four probes are combined, $\Omega_{m,0}=0.2898^{+0.00893}_{-0.0387}$.

The same trend is visible in the correlation between $\Omega_{m,0}$ and the model parameter $\alpha_{\rm phys}$. The shape of this contour, with a positive correlation between $\Omega_{m,0}$ and $\alpha_{\rm phys}$, can be attributed to the stability constraints in \cref{model_constr1}, as also observed in \cite{Csillag:2025gnz}.
For CC+PN, and still for CC+PN+RSD, the joint contour is elongated where a change in $\alpha_{\rm phys}$ can be offset by a change in $\Omega_{m,0}$, and the preferred value of $\alpha_{\rm phys}$ sits slightly higher, with a comparatively wide uncertainty. These datasets alone therefore have a weak constraining power on $\Omega_{m0}$.
Including the BAO measurements shortens that contour and significantly improves the constraints. Both CC+PN+BAO and CC+PN+BAO+RSD prefer a slightly lower value, and for the full combination the posterior peaks at $\alpha_{\rm phys}=0.08964$. This lies close to the lower edge of the interval allowed by the background solution, rather than in the broader range that remains open when the BAO data are omitted.

One can then straightforwardly conclude that including BAO data forces the model closer to the $\Lambda$CDM limit, as calculated from \cref{alpha_lambda} using the central values for CC+PN+BAO+RSD, giving $\alpha_{\rm phys}=0.08964$ and $\alpha_{\Lambda}\approx0.088$. Compare this to the corresponding CC+PN+RSD values, where one has $\alpha_{\rm phys}=0.09295$ and the $\Lambda$CDM limit of $\alpha_{\Lambda}\approx0.08856$. This phenomenon, of different dataset combinations bringing models closer or further from their $\Lambda$CDM limit, will be key to interpreting the overall performance of each model, as discussed in the next section.

The amplitude $\sigma_{8,0}$ appears in the figure only for the combinations that contain RSD data, since the other probes do not measure the growth of structure. For CC+PN+BAO+RSD the posterior gives $\sigma_{8,0}=0.7214^{+0.0948}_{-0.0657}$. For the full combinations the table also gives $D'(0)=-0.5322^{+0.0479}_{-0.0408}$. With the normalisation $D(0)=1$, this is simply the present-day growth rate through $D'(0)=-f(0)$, but it should be noted that
this can contain both growing and decaying modes. For the full combination we find
$M=-19.26^{+0.0292}_{-0.0275}$,  consistent with the SH0ES calibration \cite{Riess:2021jrx}. The sound horizon inferred from the BAO scale is $r_d=137.5^{+2.05}_{-1.89}\,\mathrm{Mpc}$, noticeably below the Planck 2018 value $r_d=147.09\pm0.26\,\mathrm{Mpc}$ \cite{Planck:2018vyg}. The BAO measurements constrain $D_M/r_d$ and $D_H/r_d$, and therefore the product $H_0 r_d$, rather than $r_d$ alone. The higher value of $H_0$ preferred by the chronometers and supernovae then shifts $r_d$ downward so that those ratios remain satisfied.

\begin{table}[htbp]
    \centering
    \resizebox{\textwidth}{!}{%
    \begin{tabular}{lccccccc}
\hline
Dataset & $H_0$ & $\Omega_{m,0}$ & $\alpha_{\rm phys}$ & $M$ & $r_{\rm drag}$ & $\sigma_{8,0}$ & $D'(0)$ \\
\hline
CC+PN & $73.18^{+0.983}_{-0.978}$ & $0.2817^{+0.0177}_{-0.134}$ & $0.09234^{+0.00273}_{-0.00201}$ & $-19.26^{+0.0282}_{-0.0288}$ & — & — & — \\
CC+PN+BAO & $73.51^{+0.933}_{-1.04}$ & $0.2886^{+0.00849}_{-0.0434}$ & $0.08944^{+0.00115}_{-0.00104}$ & $-19.25^{+0.0272}_{-0.0304}$ & $137.6^{+2.06}_{-1.91}$ & — & — \\
CC+PN+RSD & $73.09^{+1.08}_{-0.912}$ & $0.2943^{+0.0189}_{-0.0814}$ & $0.09295^{+0.00286}_{-0.00187}$ & $-19.26^{+0.0292}_{-0.0285}$ & — & $0.7131^{+0.171}_{-0.0653}$ & $-0.541^{+0.102}_{-0.0509}$ \\
CC+PN+BAO+RSD & $73.41^{+1}_{-0.928}$ & $0.2898^{+0.00893}_{-0.0387}$ & $0.08964^{+0.00107}_{-0.00105}$ & $-19.26^{+0.0292}_{-0.0275}$ & $137.5^{+2.05}_{-1.89}$ & $0.7214^{+0.0948}_{-0.0657}$ & $-0.5322^{+0.0479}_{-0.0408}$ \\
\hline
\end{tabular}
    }
    \caption{Constraints for Model~1. Central values are the maxima of the marginalized one-dimensional posteriors, with equal-tailed $68\%$ credible intervals. The parameters $\sigma_{8,0}$ and $D'(0)$ are reported only for combinations that include RSD data, and $r_{\rm drag}$ only for combinations that include BAO data. The supernova absolute magnitude $M$ is sampled in all combinations. Dashes indicate parameters not sampled for the corresponding data combination.}
    \label{tab:model1}
\end{table}

\subsection{Model~2 constraints}

\begin{figure}[htbp]
    \centering
    \includegraphics[width = \linewidth]{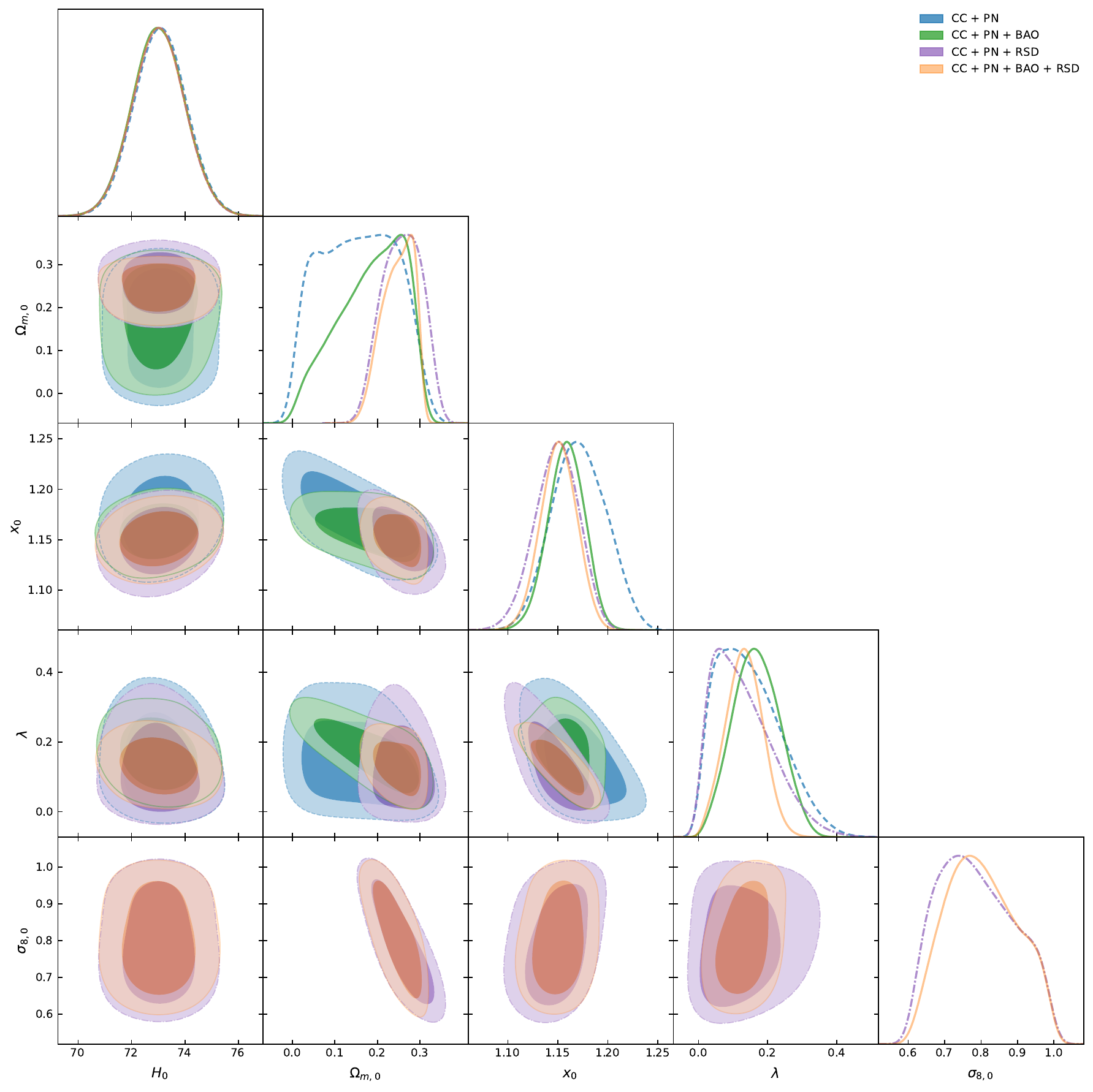}
    \caption{Confidence contours and posterior distributions for Model~2. Diagonal panels show the one-dimensional posteriors, and off-diagonal panels the corresponding $68\%$ and $95\%$ constraints, for CC+PN, CC+PN+BAO, CC+PN+RSD and CC+PN+BAO+RSD. The amplitude $\sigma_{8,0}$ is shown only for the combinations that include RSD data.}
    \label{fig:model2}
\end{figure}

The posteriors and confidence intervals for Model~2 are shown in Fig.~\ref{fig:model2}, and the corresponding constraints are given in Table~\ref{tab:model2}. As for Model~1, $H_0$ is stable across the four combinations, both in its central value and in its uncertainty. The full combination gives $H_0=73.01^{+0.988}_{-0.969}\,\mathrm{km\,s^{-1}\,Mpc^{-1}}$.

The matter density parameter is much more sensitive to
the choice of dataset. With CC+PN alone, the posterior
is broad, and adding BAO shifts the peak to higher values
without removing the long tail toward low $\Omega_{m,0}$.
Including growth data substantially tightens the constraints.
In Fig.~\ref{fig:model2}, the purple contours (CC+PN+RSD)
and the orange contours (CC+PN+BAO+RSD) are both much
narrower than those obtained without RSD, and the full
dataset combination gives
$\Omega_{m,0}=0.2791^{+0.00792}_{-0.0711}$.
Relative to CC+PN+BAO, including RSD reduces the width
of the $68\%$ credible interval for $\Omega_{m,0}$
by a factor of approximately $2.2$.
The corresponding $\Omega_{m,0}$--$\sigma_{8,0}$ contours
show the expected anticorrelation: a larger matter density
enhances growth, favouring a lower present-day amplitude
$\sigma_{8,0}$ to reproduce the same $f\sigma_8$ measurements.

The two parameters that define the scalar field behave differently from one another. The quantity $x_0$ is the present value of the dimensionless field variable. Together with $\Omega_{m,0}$ it fixes the second field variable through the constraint equation, and therefore fixes how the dark-energy sector contributes to $H(z)$ today. Distance information is already enough to measure it. All four combinations prefer nearly the same value, and the full combination gives $x_0=1.151^{+0.0179}_{-0.0182}$, only slightly lower than the CC+PN result.

The parameter $\lambda$ controls the exponential dependence of the quadratic kinetic term on the scalar field, with $\lambda=0$ recovering Model~1. CC+PN leaves this direction open, in which the posterior peaks at a low value but has a long tail toward larger $\lambda$ values, and adding RSD data does not considerably remove that tail. BAO measurements probe the expansion at several redshifts
and are therefore sensitive to the changes in the expansion
history produced by $\lambda$. The posterior then shifts upward and becomes much more symmetric, and the full combination provides the tightest result, $\lambda=0.1319^{+0.0537}_{-0.0543}$. What is particularly physically interesting is that the inclusion of BAO data increases the marginal posterior peak of $\lambda$, pushing the model further from $\Lambda$CDM and Model~1. This is in contrast to the effects of BAO data in Model~1, which has the opposite effect.

The amplitude $\sigma_{8,0}$ is constrained only when the RSD data are included. The CC+PN+RSD posterior is broad, while the full combination tightens it to $\sigma_{8,0}=0.7688^{+0.145}_{-0.0676}$. For the same combinations, $D'(0)=-0.5097^{+0.084}_{-0.0361}$. The supernova absolute magnitude remains $M=-19.26^{+0.0289}_{-0.0282}$, consistent with the SH0ES calibration $M=-19.253\pm0.029$. The sound horizon is $r_d=136.6^{+2.06}_{-1.95}\,\mathrm{Mpc}$, again below the Planck 2018 value $r_d=147.09\pm0.26\,\mathrm{Mpc}$, for the same reasons as previously discussed.

\begin{table}[htbp]
    \centering
    \resizebox{\textwidth}{!}{%
    \begin{tabular}{lcccccccc}
\hline
Dataset & $H_0$ & $\Omega_{m,0}$ & $x_0$ & $\lambda$ & $M$ & $r_{\rm drag}$ & $\sigma_{8,0}$ & $D'(0)$ \\
\hline
CC+PN & $73.06^{+0.993}_{-0.964}$ & $0.2115^{+0.0402}_{-0.16}$ & $1.169^{+0.0308}_{-0.0248}$ & $0.09459^{+0.148}_{-0.0473}$ & $-19.25^{+0.0288}_{-0.028}$ & — & — & — \\
CC+PN+BAO & $72.98^{+1.02}_{-0.937}$ & $0.2558^{+0.0136}_{-0.159}$ & $1.159^{+0.0177}_{-0.0189}$ & $0.1604^{+0.0755}_{-0.0621}$ & $-19.25^{+0.0295}_{-0.0274}$ & $136.7^{+2.01}_{-1.99}$ & — & — \\
CC+PN+RSD & $73.02^{+0.972}_{-0.977}$ & $0.2696^{+0.0369}_{-0.0608}$ & $1.151^{+0.0201}_{-0.0242}$ & $0.05995^{+0.158}_{-0.0222}$ & $-19.25^{+0.028}_{-0.0287}$ & — & $0.7389^{+0.17}_{-0.0623}$ & $-0.5091^{+0.0816}_{-0.0618}$ \\
CC+PN+BAO+RSD & $73.01^{+0.988}_{-0.969}$ & $0.2791^{+0.00792}_{-0.0711}$ & $1.151^{+0.0179}_{-0.0182}$ & $0.1319^{+0.0537}_{-0.0543}$ & $-19.26^{+0.0289}_{-0.0282}$ & $136.6^{+2.06}_{-1.95}$ & $0.7688^{+0.145}_{-0.0676}$ & $-0.5097^{+0.084}_{-0.0361}$ \\
\hline
\end{tabular}
    }
    \caption{Constraints for Model~2. Central values are the maxima of the marginalized one-dimensional posteriors, with equal-tailed $68\%$ credible intervals. The parameters
$\sigma_{8,0}$ and $D'(0)$ are reported only for combinations
that include RSD data, and $r_{\rm drag}$ only for combinations
that include BAO data. The supernova absolute magnitude $M$
is sampled in all combinations. Dashes indicate parameters
not sampled for the corresponding data combination.}
    \label{tab:model2}
\end{table}

\subsection{Model comparison}
\label{sec:model_comparison}

\begin{figure}[htbp]
    \centering
    \includegraphics[width=\linewidth]{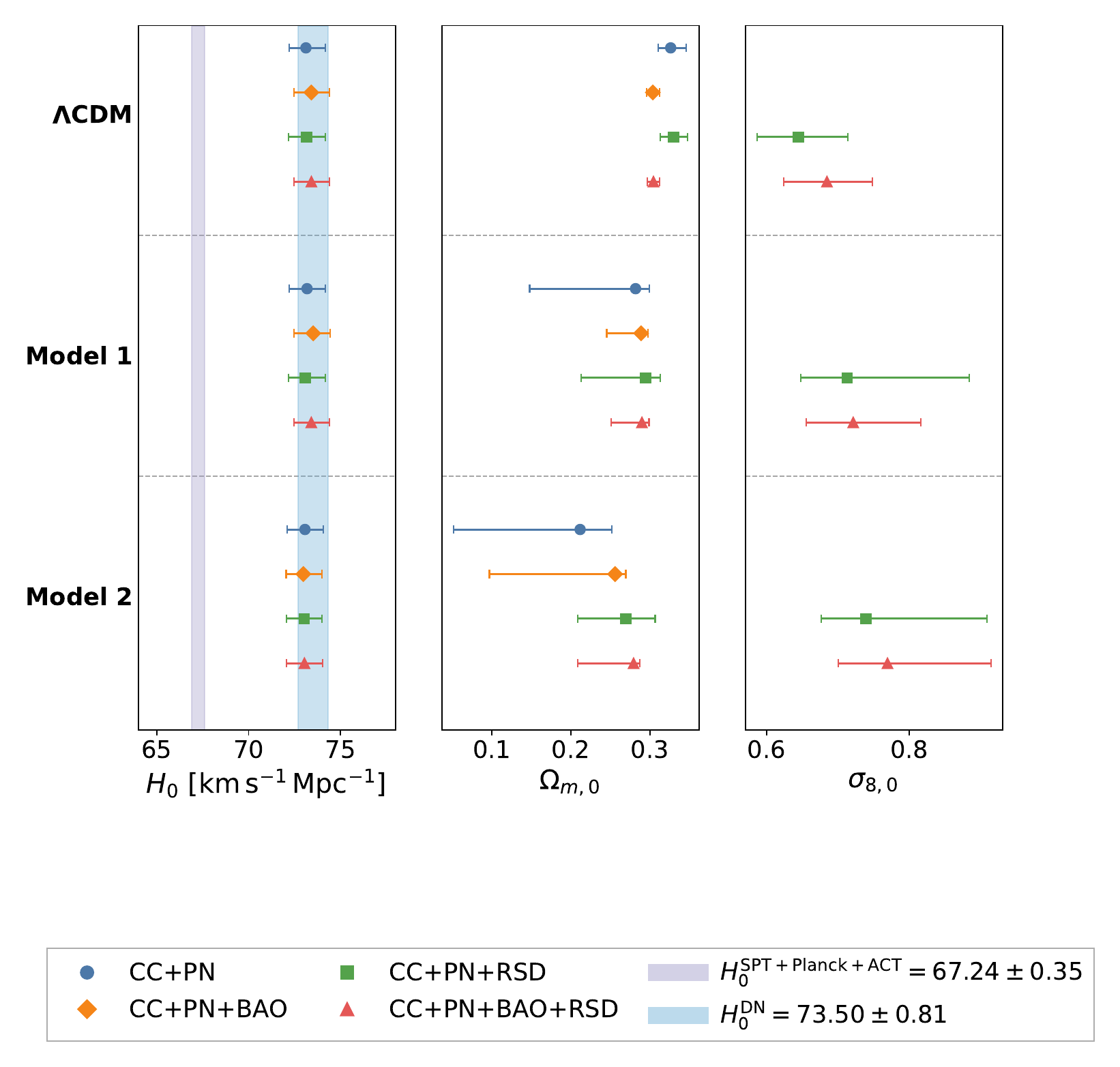}
    \caption{Constraints on $H_0$, $\Omega_{m,0}$, $\sigma_{8,0}$ and $S_8$ for $\Lambda$CDM, Model~1 and Model~2. Markers distinguish the data combinations, and bars show equal-tailed $68\%$ uncertainties. The $H_0$ reference bands show the combined SPT--Planck--ACT result
and the distance-network determination.}
    \label{fig:whisker_plot}
\end{figure}

Figure~\ref{fig:whisker_plot} brings the three models together. Each row is one data combination within a specified model, and the horizontal bars are the equal-tailed $68\%$ uncertainties. Shaded bands mark the reference determinations: the combined SPT--Planck--ACT CMB result, $H_0=67.24\pm0.35\,\mathrm{km\,s^{-1}\,Mpc^{-1}}$ \cite{SPT-3G:2025bzu}, and the distance network, $H_0=73.50\pm0.81\,\mathrm{km\,s^{-1}\,Mpc^{-1}}$ \cite{H0DN:2025lyy}, in the Hubble panel, and KiDS-Legacy \cite{Wright:2025xka}, DES Y6 \cite{DES:2026fyc}
and Planck 2018 \cite{Planck:2018vyg} in the $S_8$ panel.

With the SH0ES-calibrated Pantheon+ sample, our $H_0$ constraints are consistent with the local distance-network determination. Across all three models and four data combinations, the central
values range from $72.98$ to $73.51\,\mathrm{km\,s^{-1}\,Mpc^{-1}}$, all within the quoted $1\sigma$ interval of the distance network. Changing the model or adding BAO or RSD shifts the central value by less than its uncertainty. This agreement with local determinations of $H_0$ is somewhat expected when using late-time data alone. Consequently, all combinations remain in tension
with the combined SPT--Planck--ACT inference.

The matter density parameter is more sensitive to the choice of
dataset. For $\Lambda$CDM, CC+PN and CC+PN+RSD both give
$\Omega_{m,0}\simeq0.33$, while the combinations including BAO
give $\Omega_{m,0}\simeq0.30$. For Model~1, BAO measurements
substantially tighten the constraint, with the full combination
giving $\Omega_{m,0}=0.2898^{+0.00893}_{-0.0387}$. For Model~2,
adding BAO to CC+PN shifts the posterior peak from $0.2115$ to
$0.2558$, but leaves a long tail towards low matter densities.
Including RSD tightens this constraint to
$\Omega_{m,0}=0.2791^{+0.00792}_{-0.0711}$ for the full
combination. The central values are therefore closer for the full
combination, with $\Lambda$CDM still preferring a slightly higher
matter density than either $k$-essence model.

The inclusion of RSD also constrains the present-day amplitude $\sigma_{8,0}$, shown in
Fig.~\ref{fig:whisker_plot}. However, since we marginalise over the slope parameter $D'(0)$ rather than assuming standard growing-mode initial conditions, the values of  $\sigma_{8,0}$ depend on the allowed mixture of growing and decaying modes. The constraints should therefore be interpreted phenomenologically, as opposed to early-time theoretically-motivated predictions. A more complete investigation of their clustering properties would require a full perturbation analysis. Nonetheless, even with the additional freedom of sampling both $D'(0)$ and $\sigma_{8,0}$, we see that including RSD data tightens the constraints on the other parameters and
breaks the $\Omega_{m,0}$ degeneracy in Model 2.

\begin{figure}[htbp]
    \centering
    \includegraphics[trim=0 0 0 18, clip, width=0.8\linewidth]{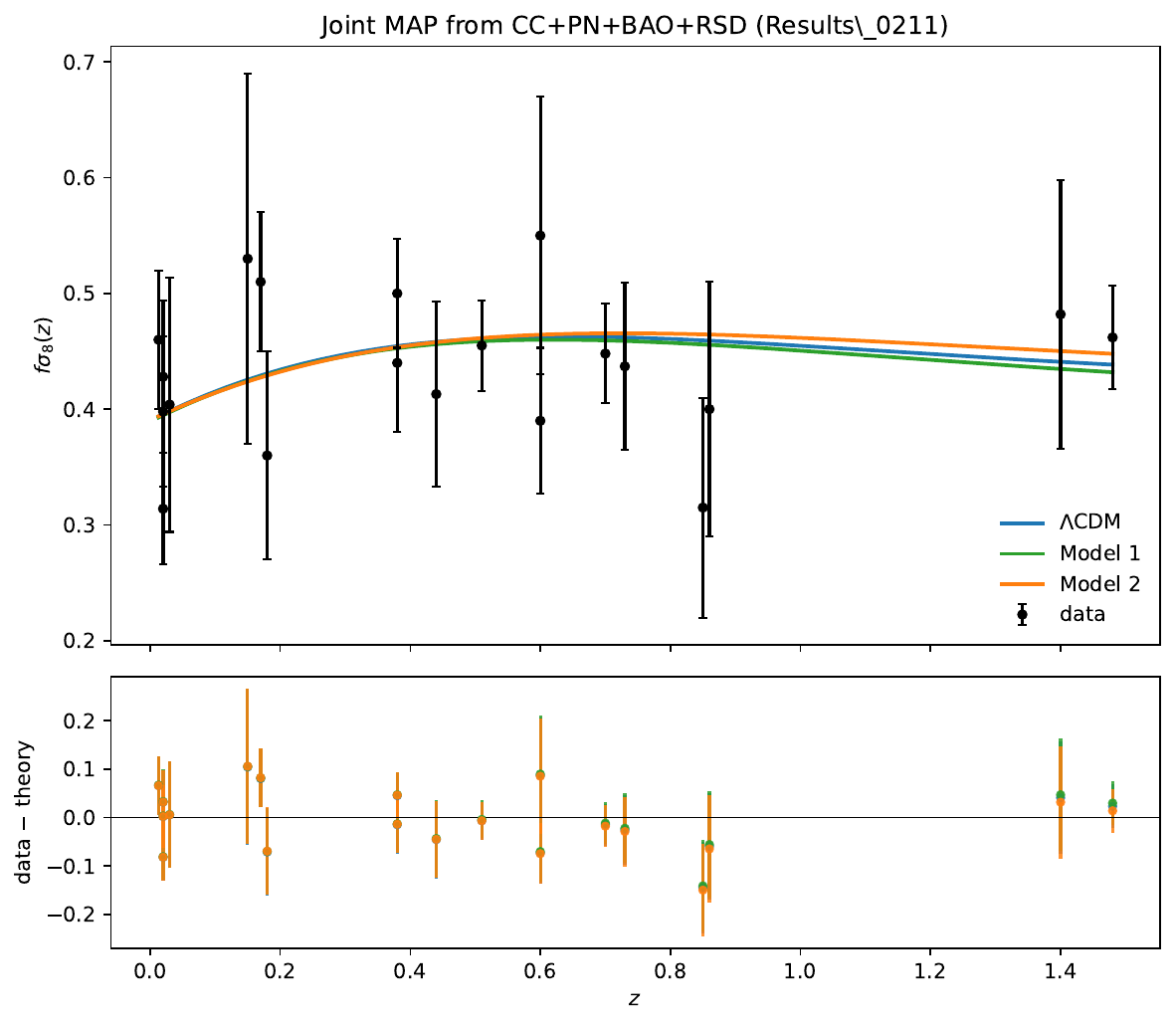}
    \caption{Best-fit $f\sigma_8(z)$ for $\Lambda$CDM, Model~1 and Model~2 from the CC+PN+BAO+RSD chains, compared with the RSD measurements. The lower panel shows the residuals (data minus model). }
    \label{fig:fs8}
\end{figure}

Figure~\ref{fig:fs8} compares the best-fit predictions
for $f\sigma_8(z)$ from the full data combination with
the RSD measurements. The three models give nearly
indistinguishable predictions over the observed redshift
range, with differences small compared with the
observational uncertainties. The lower panel shows
the corresponding residuals.

To assess the performance of each model, for each corresponding dataset combination, we compare the minimum chi-squared and Akaike and Bayesian information criteria. The priors are flat inside the sampled region, so the sample with the highest log-probability also has the highest likelihood, giving
\begin{equation}
    \chi^2_{\min} = -2\ln L_{\max}\,.
\end{equation}
A smaller value means a closer fit, before any penalty for the number of free parameters.

The parameter count is not the same for every combination. The absolute magnitude $M$ is free whenever PN is included, $r_d$ only when BAO data are included, and $\sigma_{8,0}$ together with $D'(0)$ only when RSD data is included. Relative to $\Lambda$CDM, Model~1 adds $\alpha_{\rm phys}$ and Model~2 adds $x_0$ and $\lambda$. Writing $l$ for the number of parameters varied in a given fit, the information criteria are
\begin{align}
    \mathrm{AIC} &= \chi^2_{\min} + 2l\,,\\
    \mathrm{BIC} &= \chi^2_{\min} + l\ln N\,,
\end{align}
where $N$ is the number of data points entering that likelihood. After all selection cuts, the number of measurements entering the
likelihoods are $N_{\rm CC}=15$, $N_{\rm PN}=1657$,
$N_{\rm BAO}=13$ and $N_{\rm RSD}=20$.

Both criteria penalise additional parameters. The BIC penalty grows with $N$, so it is the more conservative of the two for these large samples. Models are compared with $\Lambda$CDM at fixed data combination through
\begin{equation}
    \Delta\chi^2 = \chi^2_{\min}-\chi^2_{\min,\Lambda\mathrm{CDM}}\,,
    \qquad
    \Delta\mathrm{AIC} = \mathrm{AIC}-\mathrm{AIC}_{\Lambda\mathrm{CDM}}\,,
    \qquad
    \Delta\mathrm{BIC} = \mathrm{BIC}-\mathrm{BIC}_{\Lambda\mathrm{CDM}}\,.
\end{equation}
A negative $\Delta\chi^2$ means a closer fit than $\Lambda$CDM. A positive $\Delta\mathrm{AIC}$ or $\Delta\mathrm{BIC}$ means that, once the extra parameters are penalised, $\Lambda$CDM is preferred for that combination. The corresponding $\Lambda$CDM constraints, together with the values of $\chi^2_{\min}$, AIC and BIC from which the differences are computed, are reported in Appendix~\ref{sec:lcdm}. 

For Model~1, $\Delta\chi^2$ is indistinguishable from zero for every data combination, so allowing $\alpha_{\rm phys}$
to vary does not improve the fit relative to $\Lambda$CDM.
The resulting $\Delta{\rm AIC}\simeq2$ is therefore almost
entirely the penalty for the additional parameter. Model~2
behaves differently when BAO data are included:
$\Delta\chi^2\simeq-4.6$ for CC+PN+BAO and
$\Delta\chi^2\simeq-4.1$ for the full combination.
This improvement approximately compensates the AIC penalty
for its two additional parameters, giving
$\Delta{\rm AIC}=-0.62$ and $-0.10$, respectively. Thus the
AIC gives at most a marginal preference for Model~2, with
essentially no distinction from $\Lambda$CDM for the full
data combination.

The BIC gives a stronger preference for $\Lambda$CDM because
of its larger penalty for additional parameters in the
large data sample. Model~1 has $\Delta{\rm BIC}\simeq7.43$--$7.56$, while Model~2 gives
$\Delta{\rm BIC}\simeq10.24$--$14.77$, depending on the combination. Thus neither extension
is preferred by the BIC. Nevertheless, at the level of the maximum likelihood
both models fit the observations comparably to
$\Lambda$CDM, with Model~2 providing a moderately
better raw fit when BAO data are included.

\begin{table}[htbp]
\centering
\begin{tabular}{llcccccc}
\hline
Model & Dataset & $\chi^2_{\min}$ & AIC & BIC & $\Delta\chi^2$ & $\Delta$AIC & $\Delta$BIC \\
\hline
\multirow{4}{*}{Model~1}
 & CC+PN & 1460.57 & 1468.57 & 1490.26 & $+0.007$ & $+2.01$ & $+7.43$ \\
 & CC+PN+BAO & 1473.35 & 1483.35 & 1510.50 & $+0.008$ & $+2.01$ & $+7.44$ \\
 & CC+PN+RSD & 1474.51 & 1486.51 & 1519.11 & $+0.081$ & $+2.08$ & $+7.51$ \\
 & CC+PN+BAO+RSD & 1487.69 & 1501.69 & 1539.78 & $+0.118$ & $+2.12$ & $+7.56$ \\
\hline
\multirow{4}{*}{Model~2}
 & CC+PN & 1460.22 & 1470.22 & 1497.33 & $-0.346$ & $+3.65$ & $+14.50$ \\
 & CC+PN+BAO & 1468.72 & 1480.72 & 1513.30 & $-4.616$ & $-0.62$ & $+10.24$ \\
 & CC+PN+RSD & 1474.33 & 1488.33 & 1526.36 & $-0.102$ & $+3.90$ & $+14.77$ \\
 & CC+PN+BAO+RSD & 1483.48 & 1499.48 & 1543.01 & $-4.099$ & $-0.10$ & $+10.78$ \\
\hline
\end{tabular}
\caption{Minimum $\chi^2$, AIC and BIC for Model~1 and Model~2. $\Delta\chi^2$, $\Delta$AIC and $\Delta$BIC are relative to $\Lambda$CDM for the same dataset ($\Delta X=X_{\mathrm{model}}-X_{\Lambda\mathrm{CDM}}$; $N=1672/1685/1692/1705$ respectively). Negative values favour the alternative model.}
\label{tab:AIC_BIC}
\end{table}

\section{Conclusions}

In this paper we have demonstrated that including RSD data can break the degeneracies present in dynamical dark energy models, focussing on models within the $k$-essence class.
In particular, the addition of RSD data has only a modest effect on the ghost condensate (Model~1) constraints once BAO measurements are included, but significantly tightens the constraints on the dilatonic ghost condensate (Model~2).
Since growth-rate data primarily probes matter clustering in the late Universe, their effects are most pronounced in constraining $\Omega_{m,0}$. Nonetheless, the marginal posterior peaks for the other background parameters were mostly unchanged between the datasets with and without RSD measurements. On the other hand, BAO data had a more pronounced effect on the marginal posterior peaks: in Model~1, including BAO data led to a visible decrease in $\alpha_{\rm phys}$, as observed in \cref{fig:model1,tab:model1}, pushing the model closer to the $\Lambda$CDM limit. In Model~2, the effect is to increase $\lambda$, which has the opposite effect by moving the equation of state of the late-time accelerating fixed point further from $w_{\phi}=-1$ \cite{Gumjudpai:2005ry}. Including BAO data also gives Model~2 a lower minimum $\chi^2$ than $\Lambda$CDM and Model~1. However, while the AIC differs little between Model~2 and $\Lambda$CDM, the BIC
favours $\Lambda$CDM throughout due to the additional model complexity.

An important part of this analysis is the restriction to physically admissible cosmological solutions. We impose the absence of ghosts and gradient instabilities within the MCMC analysis, automatically rejecting non-physical samples. For Model~2, this requires checking the stability conditions along each background trajectory, since stable initial conditions alone do not ensure a stable evolution. We also explicitly examine the conditions for the quasi-static approximation on the relevant RSD scales, providing a quantitative basis for the growth-rate analysis.

Our results strongly motivate the inclusion of growth-rate
data for models which present background degeneracies,
especially those involving the matter density parameter
$\Omega_{m,0}$. Since the scalar field clusters differently
from standard pressureless matter, RSD measurements can
distinguish contributions that are degenerate in the
background expansion. This is particularly relevant to
previous background analyses of related $k$-essence models,
where $\Omega_{m,0}$ remains weakly constrained
\cite{Hussain:2024qrd}. Parameter degeneracies also arise
in unified $k$-essence models, where the kinetic scale
$X_0$ was fixed in the perturbation analysis because it
remained poorly constrained by the data
\cite{Moreno:2025wki}. Combining growth-rate measurements
with theoretical stability requirements and standard
background probes therefore provides a fast and efficient
way to constrain potentially degenerate dark sector models.

Future work should be directed toward better understanding
the observational viability of the ghost and dilatonic
ghost condensate models, as well as suitable extensions.
For instance, couplings between the dilatonic ghost
condensate and dark matter have been studied for their
ability to produce scaling solutions and modify the
late-time evolution
\cite{Gumjudpai:2005ry,Gonzalez-Espinoza:2025bzd}.
Such interactions also modify matter growth, making RSD
measurements a natural complement to existing background
analyses. However, extending our analysis to these models
requires deriving the corresponding perturbative stability
conditions and growth equations, as well as checking the
validity of the quasi-static approximation and the relation
between matter growth and the velocities probed by RSD. Nonetheless, since these extended models will likely have more significant deviations from $\Lambda$CDM, testing their observational viability is the next important step, with growth-rate data being a key tool well suited for this purpose.

\begingroup
\makeatletter
\def\addcontentsline#1#2#3{}
\acknowledgments
\makeatother
This paper is based upon work from COST Action CA21136 Addressing observational tensions in cosmology with systematics and fundamental physics (CosmoVerse).
EJ is supported by the Engineering and Physical Sciences Research Council (EPSRC) [EP/W524335/1, UKRI3030]. 
R.B. would like to acknowledge funding from “The Malta Council for Science and Technology”
as part of the “Technology Development Programme” DTP-2024-014
(CosmicLearning)
Project. 
\endgroup

\appendix
\crefalias{section}{appendix}

\section{Parameter reduction for Model~2} \label{append:calc}

For the dilatonic ghost condensate model, the Friedmann and acceleration equations are
\begin{align}
3H^2 &=  \rho
- \frac{1}{2}\dot{\phi}^2
+ \frac{3}{4} \beta e^{\lambda \phi} \dot{\phi}^4 \, , \\ 
2\dot{H} + 3H^2&= - p
+ \frac{1}{2}\dot{\phi}^2
- \frac{1}{4} \beta e^{\lambda \phi} \dot{\phi}^4 \, ,
\end{align}
while the scalar field equation is 
\begin{equation}
\left(1-3\beta e^{\lambda\phi}\dot{\phi}^2\right)\ddot{\phi}
+
3H\left(1-\beta e^{\lambda\phi}\dot{\phi}^2\right)\dot{\phi}
-
\frac{3}{4}\beta\lambda e^{\lambda\phi}\dot{\phi}^4
=
0 \, .
\end{equation}
We introduce the change of variables \cite{Gumjudpai:2005ry}
\begin{equation} \label{xy}
    x=\frac{\dot{\phi}}{\sqrt{6}H}\,,
    \qquad
    y=\frac{e^{-\lambda\phi/2}}{\sqrt{3\beta} H}\,,
\end{equation}
with $x > 0$, $y >  0$, $\beta >0$, $\lambda \geq 0$ and $H>0$, as justified in the main text. These definitions imply the substitutions
\begin{equation}
    \dot{\phi}=\sqrt{6}\,xH\,,
    \qquad
    \phi=\frac{2}{\lambda}\log \left(\frac{1}{\sqrt{3\beta} H y}\right)\,.
\end{equation}
The Friedmann constraint and acceleration equation can then be written as
\begin{align} \label{fff1}
    1+x^2-\frac{3x^4}{y^2}-\frac{\rho}{3H^2} &=0\,. \\
     p+H^2\left(3-3x^2+\frac{3x^4}{y^2}\right)+2\dot{H}&=0\,. \label{fff2}
\end{align}
The scalar-field equation becomes an evolution equation for $x$,
\begin{align} \label{ff3}
    \frac{1}{y^2}\Bigg[H^2\left(-6\sqrt{6}\,x^3-9\lambda x^4+3\sqrt{6}xy^2\right) +\sqrt{6}x\left(-6x^2+y^2\right)\dot{H} +\sqrt{6}H\left(-6x^2+y^2\right)\dot{x} \Bigg] = 0 \,.
\end{align}
while the evolution equation for $y$ follows from the definitions~(\ref{xy}),
\begin{equation} \label{ff4}
    \dot{y}=-\frac{y\left(\sqrt{6}\,\lambda H^2x+2\dot{H}\right)}{2H}\,.
\end{equation}

Transforming now to redshift representation $1+z=1/a$ and using $\frac{d}{dt}=-(1+z)H\frac{d}{dz}$,
it follows that
\begin{equation}
    \dot{H}=-H_0^2(1+z)h\frac{dh}{dz}\,,\qquad
    \dot{x}=-H_0(1+z)h\frac{dx}{dz} \,,\qquad
    \dot{y}=-H_0(1+z)h\frac{dy}{dz} \,.
\end{equation}
In terms of redshift variables, the Friedmann equation becomes
\begin{equation}
    -3\Omega_{m0}(1+z)^3+h^2\left(3+3x^2-\frac{9x^4}{y^2}\right)=0\,,
\end{equation}
and the acceleration equation is
\begin{equation}
    H_0^2 h^2 \left(3-3x^2+\frac{3x^4}{y^2}\right) -2H_0^2(1+z)h\frac{dh}{dz} =0 \,.
\end{equation}
The $x$ evolution equation is then
\begin{align}
    \frac{H_0^2h}{y^2}\Bigg[\sqrt{6}(1+z)x(6x^2-y^2)\frac{dh}{dz}-h\left(6\sqrt{6}x^3 +9\lambda x^4 -3\sqrt{6}xy^2 -6\sqrt{6}(1+z)x^2\frac{dx}{dz} +\sqrt{6}(1+z)y^2\frac{dx}{dz}\right) \Bigg] =0\,,
\end{align}
and the $y$-equation is
\begin{equation}
    \frac{1}{2}H_0y\left(\sqrt{6}\lambda hx-2(1+z)\frac{dh}{dz}\right) = H_0(1+z)h\frac{dy}{dz}\,.
\end{equation}

Evaluating the Friedmann constraint at $z=0$, with $h(0)=1$, $x(0)=x_0$, and $y(0)=y_0$, gives
\begin{equation}
    3+3x_0^2-\frac{9x_0^4}{y_0^2}-3\Omega_{m0}=0\,,
\end{equation}
which can be solved for $y_0$ as
\begin{equation}
    y_0=\frac{\sqrt{3}x_0^2}{\sqrt{1+x_0^2-\Omega_{m0}}}\,,
\end{equation}
where we've taken the positive branch since $y >0$. Note that the denominator is already enforced to be positive from
 $2x^2 \geq y^2$ and $0 \leq \Omega_{m0} \leq 1$. Explicitly, we obtain constraints on the initial value of $x_0$ from these two conditions, which are
 \begin{equation}
     - \sqrt{2-2\Omega_{m0}} < x_0 < \sqrt{2-2\Omega_{m0}} \, .
 \end{equation}

The final system of first-order equations is therefore
\begin{align} \label{aa1}
    \frac{dh}{dz}
    &=
    -\frac{3h\left(-x^4-y^2+x^2y^2\right)}{2(1+z)y^2}\,,
    \\ \label{aa2}
    \frac{dx}{dz}
    &=
    -\frac{3\left(6x^7+2x^3y^2-\sqrt{6}\lambda x^4y^2-7x^5y^2+xy^4+x^3y^4\right)}{2(1+z)y^2(6x^2-y^2)}\,,
    \\ \label{aa3}
    \frac{dy}{dz}
    &=
    -\frac{3x^4+3y^2-\sqrt{6}\lambda xy^2-3x^2y^2}{2(1+z)y}\,,
\end{align}
with initial condition
\begin{equation}
    y_0=\frac{\sqrt{3}x_0^2}{\sqrt{1+x_0^2-\Omega_{m0}}}\,,
\end{equation}
as presented in \cref{sec:back2}.

\section{Validity of the quasi-static limit}
\label{append:QS}
The validity of the growth equation requires that the modes
lie well inside the Hubble horizon and that dark energy
perturbations give a subdominant contribution to the
gravitational potential \cite{Tsujikawa:2012hv}.
As discussed in the main text, the latter can occur because
pressure support suppresses perturbations on scales well
inside the sound horizon, $k\gg k_s:=aH/(c c_s)$, or
because their contribution is suppressed by the background
factor $\Omega_\phi(1+w_\phi)$ \cite{Sapone:2009mb}.
For a small sound speed, the subhorizon condition alone does not ensure the validity of the quasi-static approximation \cite{Sawicki:2015zya}.
Here we illustrate these conditions by solving the background equations \cref{FF1,Phi1} for Model~1 and
\cref{dd1,dd2,dd3} for Model~2, using representative parameter values from the analysis in \cref{sec:results}. The resulting $k$-essence solutions determine
the evolution of $H(z)$, $\Omega_\phi(z)$, $w_\phi(z)$
and $c_s^2(z)$.

\begin{figure}[htbp]
    \centering
    \includegraphics[width=\linewidth]{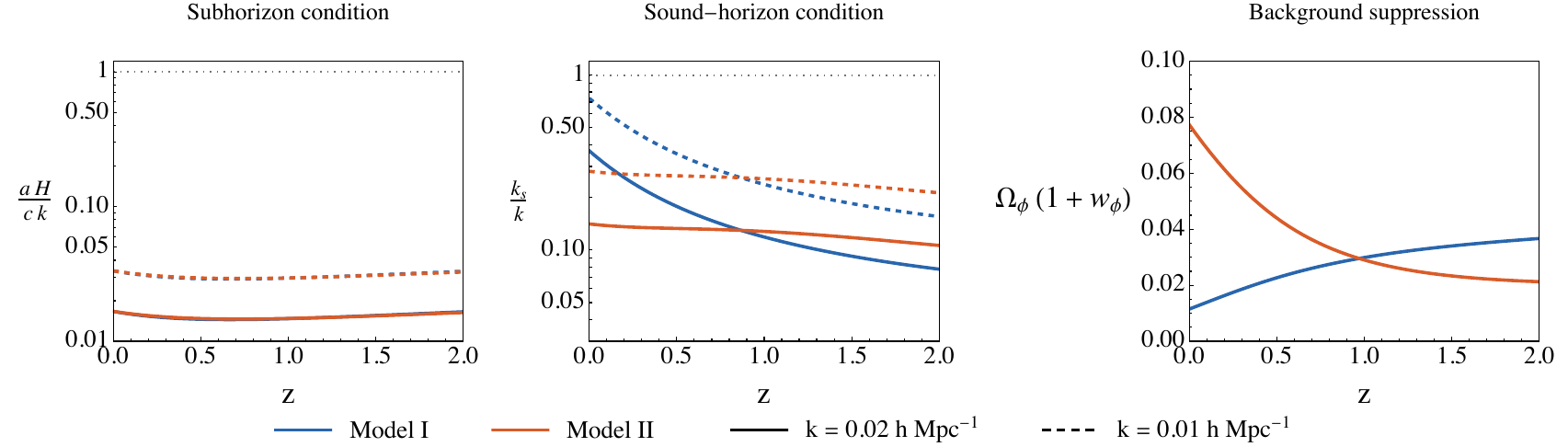}
    \caption{
    Subhorizon and sound-horizon conditions, together with
    the background suppression factor, for Models~1 (blue)
    and~2 (orange). In the first two panels, solid and
    dashed curves correspond to $k=0.02$ and
    $0.01h\,{\rm Mpc}^{-1}$, respectively, and the grey
    dotted lines mark unity. The final panel shows the
    scale-independent quantity $\Omega_\phi(1+w_\phi)$ which acts to suppress dark energy perturbations in the Poisson equation. 
    }
    \label{fig:QS}
\end{figure}

In \cref{fig:QS}, we plot $aH/(ck)$, $k_s/k$ and
$\Omega_\phi(1+w_\phi)$ over the relevant redshift range
$0\leq z\leq2$.
We consider $k=0.01$ and $0.02h\,{\rm Mpc}^{-1}$ to
illustrate the conditions on large scales. For comparison,
the 2MTF+6dFGSv measurements included in our growth
compilation use $k_{\min}=0.02h\,{\rm Mpc}^{-1}$
\cite{Qin:2019axr}, so the smaller wavenumber provides a conservative check relative to this fitting range.
Both $aH/(ck)$ and $k_s/k$ decrease inversely with $k$,
and hence the horizon conditions become better satisfied
at larger wavenumbers.
The subhorizon condition is easily satisfied, as expected,
with $aH/(ck)<0.05$.
Both models also have $k_s/k<1$ throughout this interval,
although the modes are not always well inside the sound
horizon at low redshift, particularly for Model~1.
The background factor $\Omega_\phi(1+w_\phi)$ remains
well below $0.1$ for both models, providing additional
support for neglecting the dark energy contribution to the
Poisson equation in these regimes and using \cref{eq:Poisson}.
In particular, for Model~1 it is smallest at low redshift,
where the modes are closest to the sound horizon.
These results support the qualitative argument
that dark energy perturbations are suppressed on the
scales of interest for the background solutions considered.

\section{Priors and \texorpdfstring{$\Lambda$}{lambda}CDM constraints}
\label{sec:lcdm}

The sampling priors for $\Lambda$CDM and Models~1 and~2 are listed in \cref{tab:priors}, subject to the background and stability conditions described in the main text.  For Model~1, the lower bound on
$\alpha_{\rm phys}$ follows from stability, while the upper bound is obtained by requiring accelerated expansion today, $q_0<0$ \cite{Csillag:2025gnz}.
For Model~2, we impose $\Omega_{m,0}<0.5$ for all four data combinations to improve convergence, as some walkers became trapped at high $\Omega_{m,0}$ in a region of very low likelihood. We verified that this restriction did not affect the reported constraints.

\begin{table}[H]
    \centering
    \renewcommand{\arraystretch}{1.5}
    \begin{tabular}{lccc}
    \hline
    Parameter & $\Lambda$CDM & Model~1 & Model~2 \\
    \hline
    \hline
    $H_0\,[\mathrm{km\,s^{-1}\,Mpc^{-1}}]$
      & $(50,100)$ & $(50,100)$ & $(50,100)$ \\
    $\Omega_{m,0}$
      & $(0.1,0.9)$ & $(0.1,0.9)$ & $(0,0.5)$ \\
    $\alpha_{\rm phys}$
      & 
      & $\bigl(\tfrac{1}{16(1-\Omega_{m,0})},\,
         \tfrac{4-3\Omega_{m,0}}{6(\Omega_{m,0}-2)^{2}}\bigr)$
      &  \\
    $x_0$
      &  & 
      & $\bigl(0,\sqrt{2(1-\Omega_{m,0})}\bigr)$ \\
    $\lambda$
      &  &  & $(0,0.817)$ \\
    $\sigma_{8,0}$
      & $(0.2,1.0)$ & $(0.2,1)$ & $(0.6,1)$ \\
    $D'(0)$
      & $(-1,1)$ & $(-1,1)$ & $(-1,1)$ \\
    $M$
      & $(-80,60)$ & $(-80,60)$ & $(-80,60)$ \\
    $r_{\rm d}\,[\mathrm{Mpc}]$
      & $(100,300)$ & $(100,300)$ & $(100,300)$ \\
    \hline
    \end{tabular}
    \caption{Uniform sampling priors for $\Lambda$CDM
    and Models~1 and~2. The parameters $\sigma_{8,0}$ and $D'(0)$
    are sampled only when RSD data are included, and
    $r_{\rm d}$ only when BAO data are included. The model-specific parameters $\alpha_{\rm phys}$, $x_0$ and $\lambda$.}
    \label{tab:priors}
\end{table}

\begin{figure}[H]
    \centering
    \includegraphics[width=0.8\textwidth]{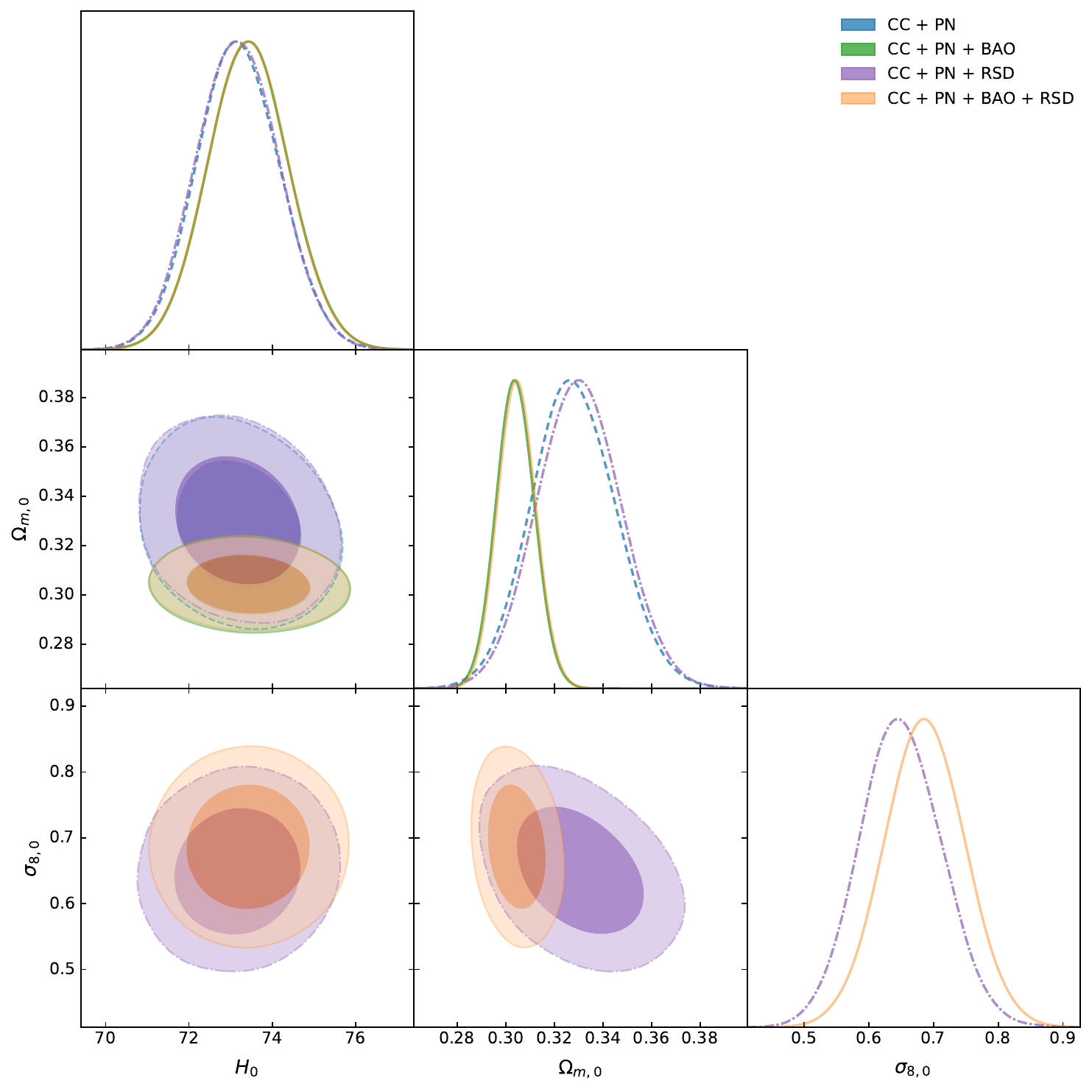}
    \caption{Confidence contours and posterior distributions for $\Lambda$CDM.}
    \label{fig:LCDM}
\end{figure}

For comparison with Models~1 and~2 in \cref{sec:model_comparison},
the $\Lambda$CDM posteriors and parameter constraints for the same
four data combinations are shown in \cref{fig:LCDM,tab:LCDM}. 
The corresponding values of $\chi^2_{\min}$, AIC and BIC are
given in \cref{tab:LCDM_AIC}.

\begin{table}[H]
    \centering
    \resizebox{\textwidth}{!}{%
    \begin{tabular}{lcccccc}
\hline
Dataset & $H_0$ & $\Omega_{m,0}$ & $M$ & $r_{\rm drag}$ & $\sigma_{8,0}$ & $D'(0)$ \\
\hline
CC+PN & $73.12^{+1.07}_{-0.917}$ & $0.3262^{+0.0194}_{-0.0154}$ & $-19.26^{+0.03}_{-0.0273}$ & — & — & — \\
CC+PN+BAO & $73.41^{+0.995}_{-0.954}$ & $0.3036^{+0.00824}_{-0.00754}$ & $-19.26^{+0.0278}_{-0.0292}$ & $137.5^{+2.06}_{-1.91}$ & — & — \\
CC+PN+RSD & $73.15^{+1.02}_{-0.973}$ & $0.3298^{+0.0177}_{-0.0169}$ & $-19.26^{+0.0285}_{-0.0291}$ & — & $0.6448^{+0.0694}_{-0.0578}$ & $-0.6016^{+0.0357}_{-0.0466}$ \\
CC+PN+BAO+RSD & $73.41^{+0.983}_{-0.963}$ & $0.3043^{+0.00793}_{-0.00772}$ & $-19.26^{+0.0281}_{-0.0289}$ & $137.5^{+2.02}_{-1.94}$ & $0.6847^{+0.0636}_{-0.0607}$ & $-0.5619^{+0.0249}_{-0.038}$ \\
\hline
\end{tabular}}
    \caption{Constraints for $\Lambda$CDM. Central values are the maxima of the marginalized one-dimensional posteriors, with equal-tailed $68\%$ credible intervals. A parameter is reported only for the combinations in which the corresponding dataset is included.}     \label{tab:LCDM}
\end{table}

\begin{table}[htbp]
\centering
\begin{tabular}{lccc}
\hline
Dataset & $\chi^2_{\min}$ & AIC & BIC \\
\hline
CC+PN & 1460.56 & 1466.56 & 1482.83 \\
CC+PN+BAO & 1473.34 & 1481.34 & 1503.06 \\
CC+PN+RSD & 1474.43 & 1484.43 & 1511.60 \\
CC+PN+BAO+RSD & 1487.58 & 1499.58 & 1532.23 \\
\hline
\end{tabular}
\caption{$\Lambda$CDM goodness-of-fit statistics.}
\label{tab:LCDM_AIC}
\end{table}

\bibliography{bib}

@article{Csillag:2025gnz,
    author = "Csillag, Lehel and Jensko, Erik",
    title = "{Geometric formulation of k-essence and late-time acceleration}",
    eprint = "2505.15975",
    archivePrefix = "arXiv",
    primaryClass = "gr-qc",
    doi = "10.1088/1475-7516/2026/09/028",
    journal = "JCAP",
    volume = "09",
    pages = "028",
    year = "2026"
}

@article{Sandvik:2002jz,
    author = "Sandvik, Havard and Tegmark, Max and Zaldarriaga, Matias and Waga, Ioav",
    title = "{The end of unified dark matter?}",
    eprint = "astro-ph/0212114",
    archivePrefix = "arXiv",
    doi = "10.1103/PhysRevD.69.123524",
    journal = "Phys. Rev. D",
    volume = "69",
    pages = "123524",
    year = "2004"
}

@article{Dubovsky:2005xd,
    author = "Dubovsky, S. and Gregoire, T. and Nicolis, A. and Rattazzi, R.",
    title = "{Null energy condition and superluminal propagation}",
    eprint = "hep-th/0512260",
    archivePrefix = "arXiv",
    reportNumber = "CERN-PH-TH-2005-265, HUTP-05-A0056",
    doi = "10.1088/1126-6708/2006/03/025",
    journal = "JHEP",
    volume = "03",
    pages = "025",
    year = "2006"
}

@article{Qin:2019axr,
    author = "Qin, Fei and Howlett, Cullan and Staveley-Smith, Lister",
    title = "{The redshift-space momentum power spectrum {\textendash} II. Measuring the growth rate from the combined 2MTF and 6dFGSv surveys}",
    eprint = "1906.02874",
    archivePrefix = "arXiv",
    primaryClass = "astro-ph.CO",
    doi = "10.1093/mnras/stz1576",
    journal = "Mon. Not. Roy. Astron. Soc.",
    volume = "487",
    number = "4",
    pages = "5235--5247",
    year = "2019"
}

@article{Giannakis:2005kr,
    author = "Giannakis, Dimitrios and Hu, Wayne",
    title = "{Kinetic unified dark matter}",
    eprint = "astro-ph/0501423",
    archivePrefix = "arXiv",
    doi = "10.1103/PhysRevD.72.063502",
    journal = "Phys. Rev. D",
    volume = "72",
    pages = "063502",
    year = "2005"
}

@article{Piazza:2004df,
    author = "Piazza, Federico and Tsujikawa, Shinji",
    title = "{Dilatonic ghost condensate as dark energy}",
    eprint = "hep-th/0405054",
    archivePrefix = "arXiv",
    reportNumber = "BICOCCA-FT-04-4",
    doi = "10.1088/1475-7516/2004/07/004",
    journal = "JCAP",
    volume = "07",
    pages = "004",
    year = "2004"
}

@article{Gumjudpai:2005ry,
    author = "Gumjudpai, Burin and Naskar, Tapan and Sami, M. and Tsujikawa, Shinji",
    title = "{Coupled dark energy: Towards a general description of the dynamics}",
    eprint = "hep-th/0502191",
    archivePrefix = "arXiv",
    doi = "10.1088/1475-7516/2005/06/007",
    journal = "JCAP",
    volume = "06",
    pages = "007",
    year = "2005"
}

@article{Tsujikawa:2012hv,
    author = "Tsujikawa, Shinji and De Felice, Antonio and Alcaniz, Jailson",
    title = "{Testing for dynamical dark energy models with redshift-space distortions}",
    eprint = "1210.4239",
    archivePrefix = "arXiv",
    primaryClass = "astro-ph.CO",
    doi = "10.1088/1475-7516/2013/01/030",
    journal = "JCAP",
    volume = "01",
    pages = "030",
    year = "2013"
}

@article{Babichev:2018twg,
    author = "Babichev, Eugeny and Ramazanov, Sabir and Vikman, Alexander",
    title = "{Recovering $P(X)$ from a canonical complex field}",
    eprint = "1807.10281",
    archivePrefix = "arXiv",
    primaryClass = "gr-qc",
    reportNumber = "LPT-Orsay-18-89",
    doi = "10.1088/1475-7516/2018/11/023",
    journal = "JCAP",
    volume = "11",
    pages = "023",
    year = "2018"
}

@article{Arkani-Hamed:2003pdi,
    author = "Arkani-Hamed, Nima and Cheng, Hsin-Chia and Luty, Markus A. and Mukohyama, Shinji",
    title = "{Ghost condensation and a consistent infrared modification of gravity}",
    eprint = "hep-th/0312099",
    archivePrefix = "arXiv",
    reportNumber = "HUTP-03-A081, UMD-PPP-04-012",
    doi = "10.1088/1126-6708/2004/05/074",
    journal = "JHEP",
    volume = "05",
    pages = "074",
    year = "2004"
}

@article{Garriga:1999vw,
    author = "Garriga, Jaume and Mukhanov, Viatcheslav F.",
    title = "{Perturbations in k-inflation}",
    eprint = "hep-th/9904176",
    archivePrefix = "arXiv",
    reportNumber = "UAB-FT-466",
    doi = "10.1016/S0370-2693(99)00602-4",
    journal = "Phys. Lett. B",
    volume = "458",
    pages = "219--225",
    year = "1999"
}

@article{Scherrer:2004au,
    author = "Scherrer, Robert J.",
    title = "{Purely kinetic k-essence as unified dark matter}",
    eprint = "astro-ph/0402316",
    archivePrefix = "arXiv",
    doi = "10.1103/PhysRevLett.93.011301",
    journal = "Phys. Rev. Lett.",
    volume = "93",
    pages = "011301",
    year = "2004"
}

@article{Armendariz-Picon:1999hyi,
    author = "Armendariz-Picon, C. and Damour, T. and Mukhanov, Viatcheslav F.",
    title = "{k - inflation}",
    eprint = "hep-th/9904075",
    archivePrefix = "arXiv",
    doi = "10.1016/S0370-2693(99)00603-6",
    journal = "Phys. Lett. B",
    volume = "458",
    pages = "209--218",
    year = "1999"
}

@article{Armendariz-Picon:2000ulo,
    author = "Armendariz-Picon, C. and Mukhanov, Viatcheslav F. and Steinhardt, Paul J.",
    title = "{Essentials of k essence}",
    eprint = "astro-ph/0006373",
    archivePrefix = "arXiv",
    doi = "10.1103/PhysRevD.63.103510",
    journal = "Phys. Rev. D",
    volume = "63",
    pages = "103510",
    year = "2001"
}

@article{Armendariz-Picon:2000nqq,
    author = "Armendariz-Picon, C. and Mukhanov, Viatcheslav F. and Steinhardt, Paul J.",
    title = "{A Dynamical solution to the problem of a small cosmological constant and late time cosmic acceleration}",
    eprint = "astro-ph/0004134",
    archivePrefix = "arXiv",
    doi = "10.1103/PhysRevLett.85.4438",
    journal = "Phys. Rev. Lett.",
    volume = "85",
    pages = "4438--4441",
    year = "2000"
}

@article{WiggleZ:2012sek,
    author = "Blake, Chris and others",
    collaboration = "WiggleZ",
    title = "{The WiggleZ Dark Energy Survey: Joint measurements of the expansion and growth history at z {\ensuremath{<}} 1}",
    eprint = "1204.3674",
    archivePrefix = "arXiv",
    primaryClass = "astro-ph.CO",
    doi = "10.1111/j.1365-2966.2012.21473.x",
    journal = "Mon. Not. Roy. Astron. Soc.",
    volume = "425",
    pages = "405--414",
    year = "2012"
}

@article{Sapone:2009mb,
    author = "Sapone, Domenico and Kunz, Martin",
    title = "{Fingerprinting Dark Energy}",
    eprint = "0909.0007",
    archivePrefix = "arXiv",
    primaryClass = "astro-ph.CO",
    doi = "10.1103/PhysRevD.80.083519",
    journal = "Phys. Rev. D",
    volume = "80",
    pages = "083519",
    year = "2009"
}

@article{Sawicki:2015zya,
    author = "Sawicki, Ignacy and Bellini, Emilio",
    title = "{Limits of quasistatic approximation in modified-gravity cosmologies}",
    eprint = "1503.06831",
    archivePrefix = "arXiv",
    primaryClass = "astro-ph.CO",
    doi = "10.1103/PhysRevD.92.084061",
    journal = "Phys. Rev. D",
    volume = "92",
    number = "8",
    pages = "084061",
    year = "2015"
}

@article{Abdalla:2022yfr,
    author = "Abdalla, Elcio and others",
    title = "{Cosmology intertwined: A review of the particle physics, astrophysics, and cosmology associated with the cosmological tensions and anomalies}",
    eprint = "2203.06142",
    archivePrefix = "arXiv",
    primaryClass = "astro-ph.CO",
    reportNumber = "FERMILAB-CONF-22-192-SCD",
    doi = "10.1016/j.jheap.2022.04.002",
    journal = "JHEAp",
    volume = "34",
    pages = "49--211",
    year = "2022"
}

@article{Verde:2019ivm,
    author = "Verde, L. and Treu, T. and Riess, A. G.",
    title = "{Tensions between the Early and the Late Universe}",
    eprint = "1907.10625",
    archivePrefix = "arXiv",
    primaryClass = "astro-ph.CO",
    doi = "10.1038/s41550-019-0902-0",
    journal = "Nature Astron.",
    volume = "3",
    pages = "891",
    year = "2019"
}

@article{Planck:2018vyg,
    author = "Aghanim, N. and others",
    collaboration = "Planck",
    title = "{Planck 2018 results. VI. Cosmological parameters}",
    eprint = "1807.06209",
    archivePrefix = "arXiv",
    primaryClass = "astro-ph.CO",
    doi = "10.1051/0004-6361/201833910",
    journal = "Astron. Astrophys.",
    volume = "641",
    pages = "A6",
    year = "2020",
    note = "[Erratum: Astron.Astrophys. 652, C4 (2021)]"
}

@article{Bamba:2011ih,
    author = "Bamba, Kazuharu and Matsumoto, Jiro and Nojiri, Shin'ichi",
    title = "{Cosmological perturbations in $k$-essence model}",
    eprint = "1109.1308",
    archivePrefix = "arXiv",
    primaryClass = "hep-th",
    doi = "10.1103/PhysRevD.85.084026",
    journal = "Phys. Rev. D",
    volume = "85",
    pages = "084026",
    year = "2012"
}

@article{dePutter:2007ny,
    author = "de Putter, Roland and Linder, Eric V.",
    title = "{Kinetic k-essence and Quintessence}",
    eprint = "0705.0400",
    archivePrefix = "arXiv",
    primaryClass = "astro-ph",
    doi = "10.1016/j.astropartphys.2007.05.011",
    journal = "Astropart. Phys.",
    volume = "28",
    pages = "263--272",
    year = "2007"
}

@article{ForemanMackey:2012ig,
    author = "Foreman-Mackey, Daniel and Hogg, David W. and Lang, Dustin and Goodman, Jonathan",
    title = "{emcee: The MCMC Hammer}",
    eprint = "1202.3665",
    archivePrefix = "arXiv",
    primaryClass = "astro-ph.IM",
    doi = "10.1086/670067",
    journal = "Publ. Astron. Soc. Pac.",
    volume = "125",
    pages = "306--312",
    year = "2013"
}

@article{Moresco:2015cya,
    author = "Moresco, Michele",
    title = "{Raising the bar: new constraints on the Hubble parameter with cosmic chronometers at z \ensuremath{\sim} 2}",
    eprint = "1503.01116",
    archivePrefix = "arXiv",
    primaryClass = "astro-ph.CO",
    doi = "10.1093/mnrasl/slv037",
    journal = "Mon. Not. Roy. Astron. Soc.",
    volume = "450",
    number = "1",
    pages = "L16--L20",
    year = "2015"
}

@article{Moresco:2012jh,
    author = "Moresco, M. and others",
    title = "{Improved constraints on the expansion rate of the Universe up to z\textasciitilde{}1.1 from the spectroscopic evolution of cosmic chronometers}",
    eprint = "1201.3609",
    archivePrefix = "arXiv",
    primaryClass = "astro-ph.CO",
    doi = "10.1088/1475-7516/2012/08/006",
    journal = "JCAP",
    volume = "08",
    pages = "006",
    year = "2012"
}

@article{Jimenez:2001gg,
      author         = "Jimenez, Raul and Loeb, Abraham",
      title          = "{Constraining cosmological parameters based on relative
                        galaxy ages}",
      journal        = "Astrophys. J.",
      volume         = "573",
      year           = "2002",
      pages          = "37-42",
      doi            = "10.1086/340549",
      eprint         = "astro-ph/0106145",
      archivePrefix  = "arXiv",
      primaryClass   = "astro-ph",
      SLACcitation   = "%%CITATION = ASTRO-PH/0106145;%%"
}

@article{Moresco:2016mzx,
    author = "Moresco, Michele and Pozzetti, Lucia and Cimatti, Andrea and Jimenez, Raul and Maraston, Claudia and Verde, Licia and Thomas, Daniel and Citro, Annalisa and Tojeiro, Rita and Wilkinson, David",
    title = "{A 6\% measurement of the Hubble parameter at $z\sim0.45$: direct evidence of the epoch of cosmic re-acceleration}",
    eprint = "1601.01701",
    archivePrefix = "arXiv",
    primaryClass = "astro-ph.CO",
    doi = "10.1088/1475-7516/2016/05/014",
    journal = "JCAP",
    volume = "05",
    pages = "014",
    year = "2016"
}

@article{Moresco:2020,
    author = "Moresco, Michele and Jimenez, Raul and Verde, Licia and Cimatti, Andrea and Pozzetti, Lucia",
    title = "{Setting the Stage for Cosmic Chronometers. II. Impact of Stellar Population Synthesis Models Systematics and Full Covariance Matrix}",
    eprint = "2003.07362",
    archivePrefix = "arXiv",
    primaryClass = "astro-ph.GA",
    doi = "10.3847/1538-4357/ab9eb0",
    journal = "Astrophys. J.",
    volume = "898",
    number = "1",
    pages = "82",
    year = "2020"
}

@article{Scolnic:2021amr,
    author = "Scolnic, Dan and others",
    title = "{The Pantheon+ Analysis: The Full Data Set and Light-curve Release}",
    eprint = "2112.03863",
    archivePrefix = "arXiv",
    primaryClass = "astro-ph.CO",
    doi = "10.3847/1538-4357/ac8b7a",
    journal = "Astrophys. J.",
    volume = "938",
    number = "2",
    pages = "113",
    year = "2022"
}

@article{Brout:2022vxf,
    author = "Brout, Dillon and others",
    title = "{The Pantheon+ Analysis: Cosmological Constraints}",
    eprint = "2202.04077",
    archivePrefix = "arXiv",
    primaryClass = "astro-ph.CO",
    doi = "10.3847/1538-4357/ac8e04",
    journal = "Astrophys. J.",
    volume = "938",
    number = "2",
    pages = "110",
    year = "2022"
}

@article{SNLS:2011lii,
    author = "Conley, A. and others",
    collaboration = "SNLS",
    title = "{Supernova Constraints and Systematic Uncertainties from the First 3 Years of the Supernova Legacy Survey}",
    eprint = "1104.1443",
    archivePrefix = "arXiv",
    primaryClass = "astro-ph.CO",
    doi = "10.1088/0067-0049/192/1/1",
    journal = "Astrophys. J. Suppl.",
    volume = "192",
    pages = "1",
    year = "2011"
}

@article{LeviSaid:2021yat,
    author = "Levi Said, Jackson and Mifsud, Jurgen and Sultana, Joseph and Adami, Kristian Zarb",
    title = "{Reconstructing teleparallel gravity with cosmic structure growth and expansion rate data}",
    eprint = "2103.05021",
    archivePrefix = "arXiv",
    primaryClass = "astro-ph.CO",
    doi = "10.1088/1475-7516/2021/06/015",
    journal = "JCAP",
    volume = "06",
    pages = "015",
    year = "2021"
}

@article{Kazantzidis:2018rnb,
    author = "Kazantzidis, Lavrentios and Perivolaropoulos, Leandros",
    title = "{Evolution of the $f\sigma_8$ tension with the Planck15/$\Lambda$CDM determination and implications for modified gravity theories}",
    eprint = "1803.01337",
    archivePrefix = "arXiv",
    primaryClass = "astro-ph.CO",
    doi = "10.1103/PhysRevD.97.103503",
    journal = "Phys. Rev. D",
    volume = "97",
    number = "10",
    pages = "103503",
    year = "2018"
}

@article{Lambiase:2018ows,
    author = "Lambiase, Gaetano and Mohanty, Subhendra and Narang, Ashish and Parashari, Priyank",
    title = "{Testing dark energy models in the light of $\sigma _8$ tension}",
    eprint = "1804.07154",
    archivePrefix = "arXiv",
    primaryClass = "astro-ph.CO",
    doi = "10.1140/epjc/s10052-019-6634-6",
    journal = "Eur. Phys. J. C",
    volume = "79",
    number = "2",
    pages = "141",
    year = "2019"
}

@article{Gonzalez:2016lur,
    author = "Gonzalez, J. E. and Alcaniz, J. S. and Carvalho, J. C.",
    title = "{Non-parametric reconstruction of cosmological matter perturbations}",
    eprint = "1602.01015",
    archivePrefix = "arXiv",
    primaryClass = "astro-ph.CO",
    doi = "10.1088/1475-7516/2016/04/016",
    journal = "JCAP",
    volume = "04",
    pages = "016",
    year = "2016"
}

@article{Gupta:2011kw,
    author = "Gupta, Gaveshna and Sen, Somasri and Sen, Anjan A",
    title = "{GCG Parametrization for Growth Function and Current Constraints}",
    eprint = "1110.0956",
    archivePrefix = "arXiv",
    primaryClass = "astro-ph.CO",
    doi = "10.1088/1475-7516/2012/04/028",
    journal = "JCAP",
    volume = "04",
    pages = "028",
    year = "2012"
}

@article{Alestas:2022gcg,
    author = "Alestas, George and Kazantzidis, Lavrentios and Nesseris, Savvas",
    title = "{Machine learning constraints on deviations from general relativity from the large scale structure of the Universe}",
    eprint = "2209.12799",
    archivePrefix = "arXiv",
    primaryClass = "astro-ph.CO",
    reportNumber = "IFT-UAM/CSIC-22-112",
    doi = "10.1103/PhysRevD.106.103519",
    journal = "Phys. Rev. D",
    volume = "106",
    number = "10",
    pages = "103519",
    year = "2022"
}

@article{Heymans:2020gsg,
    author = "Heymans, Catherine and others",
    title = "{KiDS-1000 Cosmology: Multi-probe weak gravitational lensing and spectroscopic galaxy clustering constraints}",
    eprint = "2007.15632",
    archivePrefix = "arXiv",
    primaryClass = "astro-ph.CO",
    doi = "10.1051/0004-6361/202039063",
    journal = "Astron. Astrophys.",
    volume = "646",
    pages = "A140",
    year = "2021"
}

@article{SupernovaCosmologyProject:1998vns,
    author = "Perlmutter, S. and others",
    collaboration = "Supernova Cosmology Project",
    title = "{Measurements of $\Omega$ and $\Lambda$ from 42 High Redshift Supernovae}",
    eprint = "astro-ph/9812133",
    archivePrefix = "arXiv",
    reportNumber = "LBNL-41801, LBL-41801",
    doi = "10.1086/307221",
    journal = "Astrophys. J.",
    volume = "517",
    pages = "565--586",
    year = "1999"
}

@article{SupernovaSearchTeam:1998fmf,
    author = "Riess, Adam G. and others",
    collaboration = "Supernova Search Team",
    title = "{Observational evidence from supernovae for an accelerating universe and a cosmological constant}",
    eprint = "astro-ph/9805201",
    archivePrefix = "arXiv",
    doi = "10.1086/300499",
    journal = "Astron. J.",
    volume = "116",
    pages = "1009--1038",
    year = "1998"
}

@article{Weinberg:1988cp,
    author = "Weinberg, Steven",
    editor = "Hsu, Jong-Ping and Fine, D.",
    title = "{The Cosmological Constant Problem}",
    reportNumber = "UTTG-12-88",
    doi = "10.1103/RevModPhys.61.1",
    journal = "Rev. Mod. Phys.",
    volume = "61",
    pages = "1--23",
    year = "1989"
}

@article{Martin:2012bt,
    author = "Martin, Jerome",
    title = "{Everything You Always Wanted To Know About The Cosmological Constant Problem (But Were Afraid To Ask)}",
    eprint = "1205.3365",
    archivePrefix = "arXiv",
    primaryClass = "astro-ph.CO",
    doi = "10.1016/j.crhy.2012.04.008",
    journal = "Comptes Rendus Physique",
    volume = "13",
    pages = "566--665",
    year = "2012"
}

@article{Riess:2021jrx,
    author = "Riess, Adam G. and others",
    title = "{A Comprehensive Measurement of the Local Value of the Hubble Constant with 1 km s$^{−1}$ Mpc$^{−1}$ Uncertainty from the Hubble Space Telescope and the SH0ES Team}",
    eprint = "2112.04510",
    archivePrefix = "arXiv",
    primaryClass = "astro-ph.CO",
    doi = "10.3847/2041-8213/ac5c5b",
    journal = "Astrophys. J. Lett.",
    volume = "934",
    number = "1",
    pages = "L7",
    year = "2022"
}

@article{CosmoVerseNetwork:2025alb,
    author = "Di Valentino, Eleonora and others",
    collaboration = "CosmoVerse Network",
    title = "{The CosmoVerse White Paper: Addressing observational tensions in cosmology with systematics and fundamental physics}",
    eprint = "2504.01669",
    archivePrefix = "arXiv",
    primaryClass = "astro-ph.CO",
    doi = "10.1016/j.dark.2025.101965",
    journal = "Phys. Dark Univ.",
    volume = "49",
    pages = "101965",
    year = "2025"
}

@article{Alam:2020sor,
  author = {Alam, Shadab and others},
  title = {Completed {SDSS-IV} extended Baryon Oscillation Spectroscopic Survey: Cosmological implications from two decades of spectroscopic surveys at the {Apache Point Observatory}},
  journal = {Phys. Rev. D},
  volume = {103},
  pages = {083533},
  year = {2021},
  eprint = {2007.08991},
  archivePrefix = {arXiv}
}

@article{SPT-3G:2025bzu,
    author = "Camphuis, E. and others",
    collaboration = "SPT-3G",
    title = "{SPT-3G D1: CMB temperature and polarization power spectra and cosmology from 2019 and 2020 observations of the SPT-3G main field}",
    eprint = "2506.20707",
    archivePrefix = "arXiv",
    primaryClass = "astro-ph.CO",
    reportNumber = "FERMILAB-PUB-25-0144-PPD",
    doi = "10.1103/7wt3-9v2y",
    journal = "Phys. Rev. D",
    volume = "113",
    number = "8",
    pages = "083504",
    year = "2026"
}

@article{H0DN:2025lyy,
    author = "Casertano, Stefano and others",
    collaboration = "H0DN",
    title = "{The Local Distance Network: A community consensus report on the measurement of the Hubble constant at {\ensuremath{\sim}}1{\%} precision}",
    eprint = "2510.23823",
    archivePrefix = "arXiv",
    primaryClass = "astro-ph.CO",
    doi = "10.1051/0004-6361/202557993",
    journal = "Astron. Astrophys.",
    volume = "708",
    pages = "A166",
    year = "2026"
}

@article{Wright:2025xka,
    author = "Wright, Angus H. and others",
    title = "{KiDS-Legacy: Cosmological constraints from cosmic shear with the complete Kilo-Degree Survey}",
    eprint = "2503.19441",
    archivePrefix = "arXiv",
    primaryClass = "astro-ph.CO",
    doi = "10.1051/0004-6361/202554908",
    journal = "Astron. Astrophys.",
    volume = "703",
    pages = "A158",
    year = "2025"
}

@article{DES:2026fyc,
    author = "Abbott, T. M. C. and others",
    collaboration = "DES",
    title = "{Dark Energy Survey Year 6 Results: Cosmological Constraints from Galaxy Clustering and Weak Lensing}",
    eprint = "2601.14559",
    archivePrefix = "arXiv",
    primaryClass = "astro-ph.CO",
    reportNumber = "DES-2025-0929, FERMILAB-PUB-26-0026-PPD",
    month = "1",
    year = "2026"
}

@article{Kunz:2007rk,
    author = "Kunz, Martin",
    title = "{The dark degeneracy: On the number and nature of dark components}",
    eprint = "astro-ph/0702615",
    archivePrefix = "arXiv",
    doi = "10.1103/PhysRevD.80.123001",
    journal = "Phys. Rev. D",
    volume = "80",
    pages = "123001",
    year = "2009"
}

@article{Kunz:2009yx,
    author = "Kunz, Martin and Liddle, Andrew R. and Parkinson, David and Gao, Changjun",
    title = "{Constraining the dark fluid}",
    eprint = "0908.3197",
    archivePrefix = "arXiv",
    primaryClass = "astro-ph.CO",
    doi = "10.1103/PhysRevD.80.083533",
    journal = "Phys. Rev. D",
    volume = "80",
    pages = "083533",
    year = "2009"
}

@article{Aviles:2011ak,
    author = "Aviles, Alejandro and Cervantes-Cota, Jorge L.",
    title = "{Dark degeneracy and interacting cosmic components}",
    eprint = "1108.2457",
    archivePrefix = "arXiv",
    primaryClass = "astro-ph.CO",
    doi = "10.1103/PhysRevD.84.083515",
    journal = "Phys. Rev. D",
    volume = "84",
    pages = "083515",
    year = "2011",
    note = "[Erratum: Phys.Rev.D 84, 089905 (2011)]"
}

@article{Hussain:2024qrd,
    author = "Hussain, Saddam and Nelleri, Sarath and Bhattacharya, Kaushik",
    title = "{Comprehensive study of k-essence model: dynamical system analysis and observational constraints from latest Type Ia supernova and BAO observations}",
    eprint = "2406.07179",
    archivePrefix = "arXiv",
    primaryClass = "astro-ph.CO",
    doi = "10.1088/1475-7516/2025/03/025",
    journal = "JCAP",
    volume = "03",
    pages = "025",
    year = "2025"
}

@article{Moreno:2025wki,
    author = "Moreno, Eladio and De-Santiago, Josue",
    title = "{From dark radiation to dark energy: unified cosmological evolution in K-essence models}",
    eprint = "2509.25088",
    archivePrefix = "arXiv",
    primaryClass = "astro-ph.CO",
    doi = "10.1088/1361-6382/ae456b",
    journal = "Class. Quant. Grav.",
    volume = "43",
    number = "4",
    pages = "045017",
    year = "2026"
}

@article{Ghedini:2024mdu,
    author = "Ghedini, Pietro and Hajjar, Rasmi and Mena, Olga",
    title = "{Redshift-space distortions corner interacting dark energy}",
    eprint = "2409.02700",
    archivePrefix = "arXiv",
    primaryClass = "astro-ph.CO",
    doi = "10.1016/j.dark.2024.101671",
    journal = "Phys. Dark Univ.",
    volume = "46",
    pages = "101671",
    year = "2024"
}

@article{DESI:2025zgx,
    author = "Abdul Karim, M. and others",
    collaboration = "DESI",
    title = "{DESI DR2 results. II. Measurements of baryon acoustic oscillations and cosmological constraints}",
    eprint = "2503.14738",
    archivePrefix = "arXiv",
    primaryClass = "astro-ph.CO",
    reportNumber = "FERMILAB-PUB-25-0169-PPD",
    doi = "10.1103/tr6y-kpc6",
    journal = "Phys. Rev. D",
    volume = "112",
    number = "8",
    pages = "083515",
    year = "2025"
}

@inproceedings{Hamilton:1997zq,
    author = "Hamilton, A. J. S.",
    title = "{Linear redshift distortions: A Review}",
    booktitle = "{Ringberg Workshop on Large Scale Structure}",
    eprint = "astro-ph/9708102",
    archivePrefix = "arXiv",
    doi = "10.1007/978-94-011-4960-0_17",
    month = "8",
    year = "1997"
}

@article{Gonzalez-Espinoza:2025bzd,
    author = "Gonzalez-Espinoza, Manuel and Herrera, Ram{\'o}n",
    title = "{Interacting dilatonic ghost condensate as dark energy model}",
    eprint = "2504.05435",
    archivePrefix = "arXiv",
    primaryClass = "gr-qc",
    doi = "10.1016/j.dark.2025.101976",
    journal = "Phys. Dark Univ.",
    volume = "49",
    pages = "101976",
    year = "2025"
}

@article{Goldstein:2025epp,
    author = "Goldstein, Samuel and Celoria, Marco and Schmidt, Fabian",
    title = "{Monodromic dark energy and DESI}",
    eprint = "2507.16970",
    archivePrefix = "arXiv",
    primaryClass = "astro-ph.CO",
    doi = "10.1103/396l-5txp",
    journal = "Phys. Rev. D",
    volume = "113",
    number = "2",
    pages = "023532",
    year = "2026"
}

@article{Lueiza-Colipi:2026bqc,
    author = "Lueiza-Colip{\'\i}, Andr{\'e}s and Dimakis, Nikolaos and Paliathanasis, Andronikos",
    title = "{Nontrivial Symmetries in $k$-essence Cosmology}",
    eprint = "2609.39729",
    archivePrefix = "arXiv",
    primaryClass = "gr-qc",
    month = "9",
    year = "2026"
}

@article{Lueiza-Colipi:2026gij,
    author = "Lueiza-Colip{\'\i}, Andr{\'e}s and Dimakis, Nikolaos and Leon, Genly and Paliathanasis, Andronikos",
    title = "{Symmetry-Driven $k$-Essence Cosmological Dynamics}",
    eprint = "2609.18790",
    archivePrefix = "arXiv",
    primaryClass = "gr-qc",
    month = "9",
    year = "2026"
}

\end{document}